\documentclass[prd,amsfonts,aps,nofootinbib,notitlepage,11pt,superscriptaddress]{revtex4-1}

\pdfoutput=1
\usepackage{amsmath,amssymb}
\usepackage{graphicx}
\usepackage[utf8]{inputenc}
\usepackage{cancel}
\usepackage[colorlinks=true]{hyperref}
\usepackage{color}
\usepackage{array,multirow}
\usepackage{subcaption}
\usepackage{float}

\newcommand{\UPTC}{Escuela de Física, Universidad Pedagógica y Tecnológica de Colombia,\\
Avenida Central del Norte \# 39-115, Tunja, Colombia}
\newcommand{\UdeA}{Instituto de Física, Universidad de Antioquia,\\Calle 70 \# 52-21, Apartado Aéreo 1226, Medellín, Colombia}
\newcommand{\Lapth}{LAPTh, Univ. Grenoble Alpes, USMB, CNRS, S-74940 Annecy, France}
\newcommand{\skobe}{Skobeltsyn Institute of Nuclear Physics, Moscow State University, Moscow 119992, Russia}
\newcommand{\ICTP}{Abdus Salam International Centre for Theoretical Physics,\\ Strada Costiera 11, 34151, Trieste, Italy.}

\begin{document}
\title{The  $Z_5$ model of two-component  dark matter}

\author{Genevi\`eve B\'elanger}\affiliation{\Lapth}
\author{Alexander Pukhov} \affiliation{\skobe}
\author{Carlos E. Yaguna}
\affiliation{\UPTC}
\author{\'Oscar Zapata}
\affiliation{\UdeA}\affiliation{\ICTP}

\begin{abstract}
Scenarios for multi-component scalar dark matter based on a single $Z_N$ ($N\geq 4$) symmetry are simple and well-motivated. In this paper we investigate, for the first time, the phenomenology of the $Z_5$ model for two-component dark matter. This model, which can be seen as an extension of the well-known singlet scalar model,  features two complex scalar fields--the dark matter particles--that are Standard Model singlets but have different charges under a $Z_5$ symmetry. The interactions allowed by the $Z_5$  give rise to novel processes between the dark matter particles that affect their relic densities and their detection prospects, which we study in detail. The key parameters of the model are identified and  its viable regions are characterized by means of random scans. We show that, unlike the singlet scalar model, dark matter masses below the TeV are still compatible with present data.  Even though the dark matter density turns out to be dominated by the lighter component, we find that current and future direct detection experiments may be  sensitive to signals from both dark matter particles.  
\end{abstract}

\maketitle
\section{Introduction}
Finding  the correct extension of the Standard Model (SM) that accounts for the dark matter (DM) is one of the  main open problems in fundamental physics today. Even if most of the models that have been proposed and studied implicitly assume that the observed dark matter density is explained by a single new particle, it does not have to be so~\cite{Boehm:2003ha,Ma:2006uv,Cao:2007fy,Hur:2007ur,Lee:2008pc,Zurek:2008qg,Profumo:2009tb,Baer:2011hx,Esch:2014jpa}. Scenarios in which two or more different particles contribute to the dark matter density --multi-component dark matter models-- not only are perfectly consistent with current observations but often lead to testable predictions in current and future dark matter experiments.

Among multi-component dark matter models, those featuring scalar fields that are simultaneously stabilized by  a single $Z_N$ symmetry  are particularly appealing~\cite{Batell:2010bp,Belanger:2014bga}. For $k$ dark matter particles, they require only $k$ complex scalar fields that are SM singlets but  have different charges under a $Z_N$ ($N\geq 2k$). This symmetry,  in turn,   could  be a remnant of a spontaneously broken $U(1)$ gauge symmetry and thus be related to gauge extensions of the SM. Recently, these scenarios were systematically analyzed \cite{Yaguna:2019cvp} and it was found that, surprisingly, their dark matter phenomenology has yet to be investigated in detail. With this paper, we intend to partially fill that gap. 

We  study the two-component dark matter model based on the $Z_5$ symmetry, which serves as a prototype for all the $Z_N$ scenarios in which the dark matter particles are two \emph{complex} scalars. Above all, we want to characterize the viable parameter space of this model and to determine its detection prospects. To that end, we first examine  the dark  matter  relic densities, identifying the types of processes that can modify them and the key parameters they depend on. Then,  the viable parameter space of the model is characterized by means of  random scans, which we analyze in detail. Our results indicate that the entire range of dark matter masses is allowed, that the dark matter density is always dominated by the lighter component, and that both dark matter particles may produce signals in future direct detection experiments.

The rest of the paper is organized as follows. In the next section, our notation is introduced and  the $Z_5$ model is briefly described --further details are relegated to the appendices. Section \ref{sec:pheno} is devoted to the dark matter phenomenology. In particular, the effect of the different parameters on the relic densities is elucidated. Our central results are obtained in section \ref{sec:viable}. In it, we first determine, via random scans, the viable parameter space of the model and then use it to predict its detection prospects. Section \ref{sec:disc} deals with possible extensions of our work whereas  section \ref{sec:conc} presents our conclusions. 

\section{The model}
\label{sec:model}

Let us consider a scenario with two new complex scalar fields, $\phi_{1,2}$, charged under a  $Z_5$ symmetry. The unique charge assignment (up to trivial field redefinitions) that allows both fields to be stable is \cite{Yaguna:2019cvp} 
\begin{align}
    \phi_1\sim \omega_5,\,\,\, \phi_2\sim \omega_5^2; \hspace{1cm}\omega_5=\exp(i2\pi/5). 
\end{align}
These new fields--the dark matter particles--are assumed to be  singlets of the SM gauge group whereas the SM particles are taken to be singlets under the $Z_5$.  The most general 
 $Z_5$-invariant scalar potential is then given by \cite{Belanger:2014vza}
\begin{align}\label{eq:Z5lag}
 \mathcal{V}&=\,\,-\mu^2_H|H|^2+\lambda_H|H|^4+\mu_{1}^2|\phi_1|^2+\lambda_{41}|\phi_1|^4+\lambda_{S1}|H|^2|\phi_1|^2+\mu_{2}^2|\phi_2|^2+\lambda_{42}|\phi_2|^4+\lambda_{S2}|H|^2|\phi_2|^2\nonumber\\
 & \,+\lambda_{412}|\phi_1|^2|\phi_2|^2+\frac{1}{2}\left[\mu_{S1}\phi^2_1\phi_2^{*} + \mu_{S2}\phi_2^2\phi_1 +\lambda_{31}\phi _1^3 \phi _2+\lambda_{32}\phi _1 \phi _2^{*3} + \text{H.c.}\right],
 \end{align}
where $H$ is the SM Higgs doublet.  To ensure that the model describes a two-component dark matter scenario, we assume that $\phi_{1,2}$ do not acquire a vacuum expectation value and   that their masses  satisfy $\frac{M_1}{2}<M_2<2 M_1$ so that  both are stable. In addition, due to the symmetry of the Lagrangian, we can take, without loss of generality, $\phi_2$ to be heavier than $\phi_1$ and so  $M_1<M_2<2M_1$, which is assumed from now on.  The stability conditions as well as the one-loop renormalization group equations (RGEs) for this model are given in the appendices.

Notice that the dark matter particles interact with the SM fields only through the Higgs boson. The $Z_5$ model is thus one example of the so-called Higgs-portal scenarios --see Ref. \cite{Arcadi:2019lka} for a recent review. In addition, the dark matter particles interact among themselves through trilinear and quartic interactions. The terms in brackets in equation (\ref{eq:Z5lag}) are interactions specific to the $Z_5$ symmetry we are considering, while the rest are present for any $Z_N$ (see also section \ref{sec:disc}). Had we imposed a $Z_2\times Z_2'$ instead, as often done in  two-component dark matter scenarios, all the terms in brackets would be forbidden. 

In total, this model contains $11$ new parameters ($4$ dimensionful and $7$ dimensionless), but  two of them --$\lambda_{41}$ and $\lambda_{42}$-- are irrelevant for the dark matter phenomenology and can be ignored in our analysis. The parameters $\mu_i^2$ ($i=1,2$), on the other hand, can be conveniently traded for the physical masses $M_i$ of the scalar fields,  so that the free parameters of the model may be taken to be $M_i$, $\lambda_{Si}$,  $\lambda_{412}$, $\mu_{Si}$, and $\lambda_{3i}$. The phases of $\phi_{1,2}$ can be chosen so as to make $\mu_{S1}$ and $\mu_{S2}$ real, but then $\lambda_{31}$ and $\lambda_{32}$ may be complex. In the following we will stick, for simplicity, to real parameters. Our goal is to study how these nine parameters affect the relic densities, shape the viable parameter space, and  determine the dark matter observables.   

This $Z_5$ model  can be seen as an extension of (and shares many features with) the well-known singlet scalar model~\cite{Silveira:1985rk,McDonald:1993ex,Burgess:2000yq}, which is based on the standard $Z_2$ symmetry and includes just one dark matter particle. This latter model is currently highly constrained, requiring dark matter masses right at the Higgs-resonance or above a TeV or so \cite{Cline:2013gha,Athron:2018ipf}.  We would like to know, therefore, whether this restriction on low dark matter masses still holds in the $Z_5$ model, or if the new interactions present in it weaken such bounds and allow the dark matter particles to have masses below the TeV. 

\section{Dark matter phenomenology}
\label{sec:pheno}
In this model, the dark matter particles  and  the SM particles are connected only via  Higgs-portal interactions. Thus, depending on the size of the couplings $\lambda_{Si}$, both freeze-in \cite{Hall:2009bx,Yaguna:2011qn} and  freeze-out scenarios can be envisaged for the dark matter relic densities. We will focus, in this paper, on the more compelling freeze-out realization, which has the advantage of giving rise to testable signatures in  dark matter experiments.

\subsection{The relic density}
\begin{table}[t]
    
    \begin{tabular}{c |c}
      $\phi_1$ Processes   & Type  \\
      \hline 
        $\phi_1+\phi_1^\dagger\to SM + SM$ & $1100$\\
        $\phi_1+\phi_1^\dagger \to \phi_2+\phi_2^\dagger$  & $1122$\\
        $ \phi_1^\dagger+h \to \phi_2+\phi_2 $  & $1022$\\
        $ \phi_1 + \phi_2^\dagger  \to \phi_2+\phi_2 $ & $1222$\\
        $\phi_1^\dagger + \phi_1^\dagger \to \phi_2+\phi_1  $ & $1112$\\
        $\phi_1+\phi_2 \to \phi_2^\dagger + h$ & $1220$\\
        $\phi_1+\phi_1 \to  \phi_2 + h $ & $1120$\\
    \end{tabular}\hspace{2cm}
    \begin{tabular}{c |c}
      $\phi_2$ Processes   & Type  \\
      \hline 
        $\phi_2+\phi_2^\dagger\to SM + SM$ & $2200$\\
        $\phi_2+\phi_2^\dagger \to \phi_1+\phi_1^\dagger$  & $2211$\\
        $\phi_2+\phi_2 \to \phi_1^\dagger+h$  & $2210$\\
        $\phi_2+\phi_2 \to \phi_1 + \phi_2^\dagger$ & $2212$\\
        $\phi_2+\phi_1 \to \phi_1^\dagger + \phi_1^\dagger$ & $2111$\\
        $\phi_2+\phi_1^\dagger \to \phi_1 + h$ & $2110$\\
        $ \phi_2 + h\to \phi_1+\phi_1 $ & $2011$\\
    \end{tabular}
    
    \caption{The $2\to 2$ processes that are allowed in the $Z_5$ model and that can modify the relic density of $\phi_1$ (left) and $\phi_2$ (right). $h$ denotes the SM Higgs boson. Conjugate and inverse processes are not shown.  
    }
    \label{tab:processes}
\end{table}

The full set of $2\to 2$ processes that may contribute to the relic density in an arbitrary two-component dark matter scenario was listed in Ref.~\cite{Belanger:2014vza}. They can be classified in \emph{types} that are denoted by four digits (each a $0$, $1$, or $2$) indicating the sector to which the particles involved in the process belong to --$0$ is used for SM particles,  $1$ for $\phi_1$ or $\phi_1^\dagger$, and $2$ for $\phi_2$ or $\phi_2^\dagger$. Thus, the type $2210$ includes all processes with  one SM particle and one $\phi_1$ (or  $\phi_1^\dagger$) in the final state, and with an initial state consisting of either two $\phi_2$, two $\phi_2^\dagger$, or $\phi_2$ and $\phi_2^\dagger$. Among the various types, the only ones not compatible with  the $Z_5$ symmetry are  $1110$ and $2220$. Table \ref{tab:processes} displays all the processes that contribute to the  relic densities in the $Z_5$ model, with their respective type.

\begin{figure}
\centering
\includegraphics[scale=0.9]{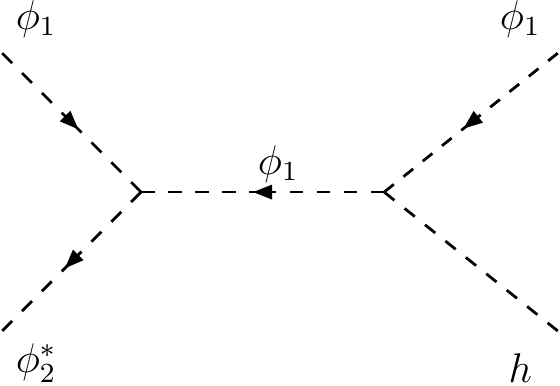}\hspace{0.4cm}
\includegraphics[scale=0.9]{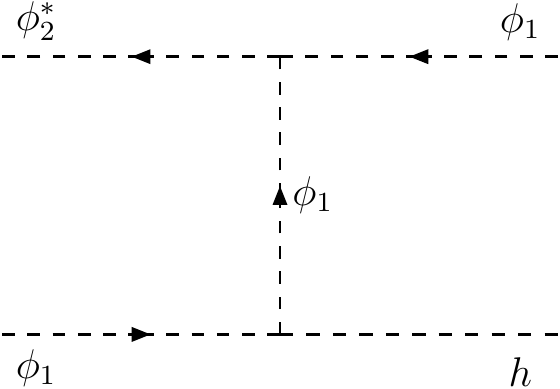}\hspace{0.4cm}
\includegraphics[scale=0.9]{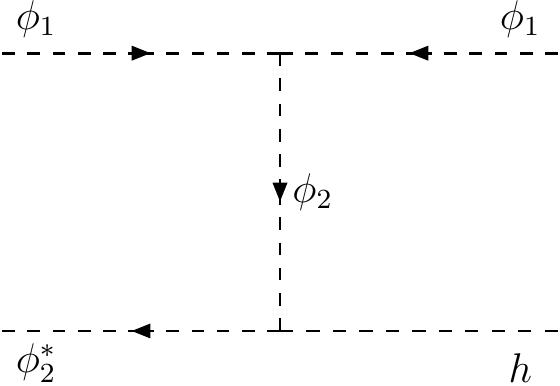}\\
\vspace{0.4cm}
\includegraphics[scale=0.9]{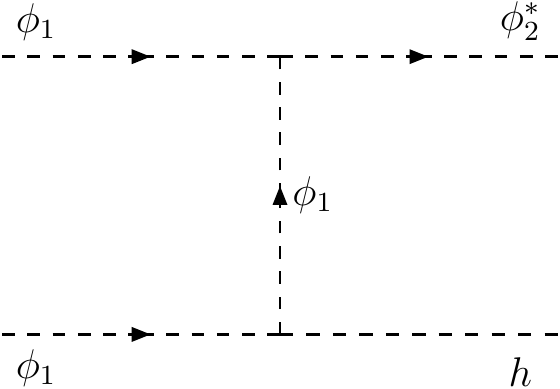}\hspace{1cm}
\includegraphics[scale=0.9]{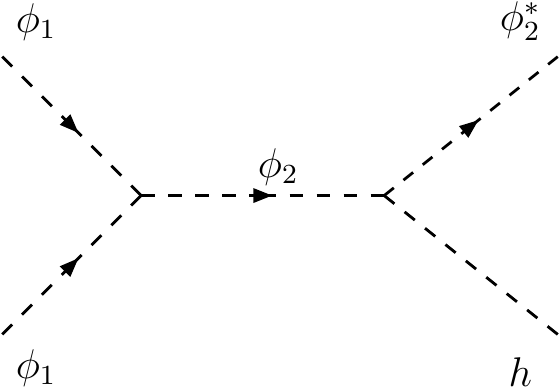}
\caption{Dark matter semi-annihilation processes involving one trilinear $\mu_{S1}$ and one Higgs-DM $\lambda_{Si}$ interactions: $\phi_1\phi_2^*\to\phi_1 h$ (top) and $\phi_2^* h\to \phi_1\phi_1$ (bottom). Replacing $\mu_{S1}\to\mu_{S2}$ similar diagrams arise for the processes $\phi_1\phi_2\to\phi_2 h$ (top) and $\phi_2\phi_2\to \phi_1 h$ (bottom).}
\label{fig:semianni2}
\end{figure}

\begin{figure}
\centering
\includegraphics[scale=0.9]{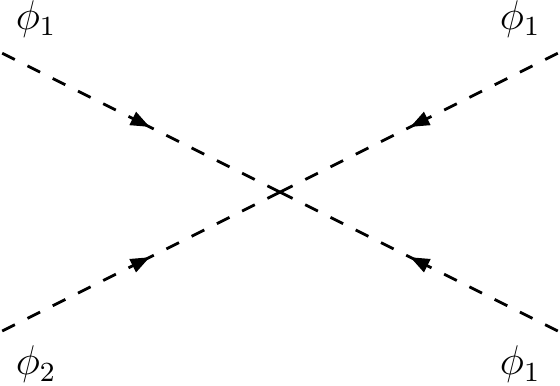}\hspace{0.4cm}
\includegraphics[scale=0.9]{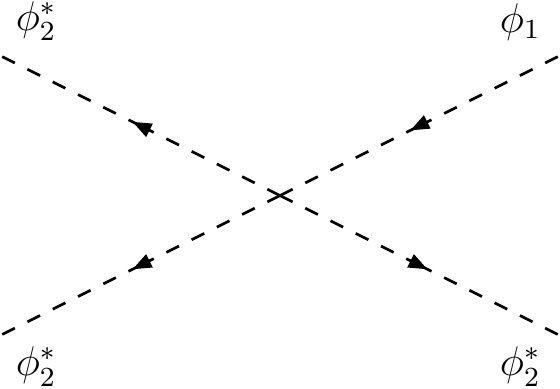}\hspace{0.4cm}
\includegraphics[scale=0.9]{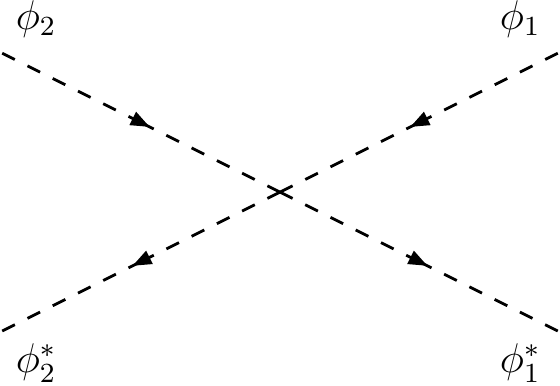}\\
\includegraphics[scale=0.9]{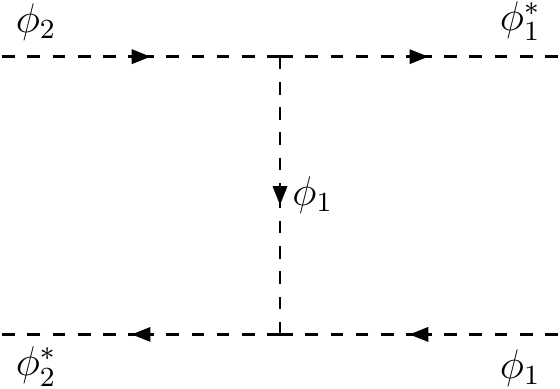}\hspace{1cm}
\includegraphics[scale=0.9]{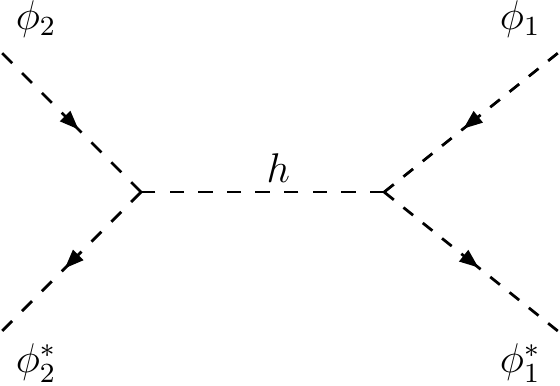}
\caption{Dark matter conversion processes. Top:  via quartic interactions --$\lambda_{31}$ (left), $\lambda_{32}$ (center) and $\lambda_{412}$ (right). Bottom:  via a $\mu_{S1}$ trilinear interaction (left) or Higgs-portal couplings (right). 
}
\label{fig:conversion2}
\end{figure}

According to the number of SM particles, these processes can be divided  into three kinds: annihilation processes (two SM particles), semi-annihilation processes \cite{DEramo:2010keq} (one SM particle), and dark matter conversion processes (no SM particles). Figures \ref{fig:semianni2} and \ref{fig:conversion2} display representative Feynman diagrams for semi-annihilation and dark matter conversion processes respectively. Given that some processes receive contributions from more than one Feynman diagram,  e.g. $\phi_2+\phi_2^\dagger\to\phi_1+\phi_1^\dagger$, interference effects are expected to play a role in certain cases. Let us also note that while the quartic couplings, $\lambda_{3i}$, induce only dark matter conversion processes, the trilinear couplings, $\mu_{Si}$, contribute to both, semi-annihilations and conversions. The annihilations into two SM particles (not shown), on the other hand,  proceed via the usual $s$-channel Higgs-mediated diagram, with $W^+W^-$  being the dominant final state for $M_i\gtrsim M_W$.

The Boltzmann equations for the $Z_5$ model thus read

\begin{eqnarray}
\label{boltzmann1}
\frac{dn_1}{dt}&=&-\sigma_v^{1100}  \left(n_1^2-\bar{n}_1^2 \right) -
\sigma_v^{1120}\left( n_1^2- n_2 \frac{\bar{n}_1^2}{\bar{n}_2} \right)
- \sigma_v^{1122}\left( n_1^2- n_2^2 \frac{\bar{n}_1^2}{\bar{n}_2^2}
\right)\nonumber  \\
  && - \frac{1}{2}\sigma_v^{1112}\left( n_1^2- n_1 n_2 \frac{\bar{n}_1}{\bar{n}_2}\right)
      - \frac{1}{2}\sigma_v^{1222}\left( n_1 n_2- n_2^2\frac{\bar{n}_1}{\bar{n}_2}\right)\nonumber \\
               &&  -\frac{1}{2}\sigma_v^{1220}\left( n_1 n_2- n_2 \bar{n}_1 \right)
+\frac{1}{2}\sigma_v^{2210}(n_2^2-n_1\frac{\bar{n}_2^2}{\bar{n}_1})   - 3H n_1\,, \\
\frac{dn_2}{dt}&=&-\sigma_v^{2200}  \left(n_2^2-\bar{n}_2^2 \right) -
\sigma_v^{2210}\left( n_2^2- n_1 \frac{\bar{n}_2^2}{\bar{n}_1} \right)
- \sigma_v^{2211}\left( n_2^2- n_1^2 \frac{\bar{n}_2^2}{\bar{n}_1^2}
\right)\nonumber  \\
  && - \frac{1}{2}\sigma_v^{2221}\left( n_2^2- n_1 n_2 \frac{\bar{n}_2}{\bar{n}_1}\right)
      - \frac{1}{2}\sigma_v^{1211}\left( n_1 n_2- n_1^2\frac{\bar{n}_2}{\bar{n}_1}\right)\nonumber \\
               &&  -\frac{1}{2}\sigma_v^{1210}\left( n_1 n_2- n_1 \bar{n}_2 \right)
+\frac{1}{2}\sigma_v^{1120}(n_1^2-n_2\frac{\bar{n}_1^2}{\bar{n}_2})   - 3H n_2.      
\label{boltzmann2}
\end{eqnarray}
Here $n_{i}$ ($i=1,2$) denote the number densities of $\phi_i$, and $\bar{n}_i$  their respective equilibrium values.  $\sigma_v^{abcd}$  stands for  the thermally averaged cross section, which satisfies 
\begin{equation}
    \bar{n}_a\bar{n}_b\sigma_v^{abcd}=\bar{n}_c\bar{n}_d\sigma_v^{cdab}.
\end{equation}
By  solving these equations, the relic densities of $\phi_1$ and $\phi_2$ --$\Omega_1$ and $\Omega_2$-- can be calculated. Since its version 4.1, {\tt micrOMEGAs}~\cite{Belanger:2014vza} incorporated two-component dark matter scenarios, automatically including all the relevant processes for a given model and numerically solving the corresponding Boltzmann equations.  It also includes the code files of the $Z_5$ model we are studying. We will rely on {\tt micrOMEGAs} \cite{Belanger:2013oya,Belanger:2014vza,Belanger:2018ccd} to compute the relic densities and the dark matter detection observables. Keep in mind, though, that in the course of this work we found and corrected some bugs affecting the calculation of the relic density for two dark matter particles. To reproduce our results, {\tt micrOMEGAs} version 5.2 or later should be used. 

To estimate the relevance of the three kinds of processes--annihilations, semi-annihilations, and conversions--that can contribute to the relic density of $\phi_1$, it is convenient to define the following three parameters
\begin{align}
    \zeta^1_{anni}&\equiv\frac{\sigma_v^{1100}}{\overline{\sigma_v^{1}}},\quad
    \zeta^1_{semi}\equiv\frac{\frac{1}{2}(\sigma_v^{1120}+\sigma_v^{1220}+\sigma_v^{1022})}{\overline{\sigma_v^{1}}},\quad
    \zeta^1_{conv}\equiv\frac{\sigma_v^{1122}+\sigma_v^{1112}+\sigma_v^{1222}}{\overline{\sigma_v^{1}}},\label{eq:semi}
\end{align}
with
\begin{align}
\overline{\sigma_v^{1}}\equiv\sigma_v^{1100}+\frac{1}{2}\sigma_v^{1120}+\sigma_v^{1122}+\sigma_v^{1112}+\sigma_v^{1222}+\frac{1}{2}\sigma_v^{1220}+\frac{1}{2}\sigma_v^{1022}.
\end{align}
These parameters are assumed to be evaluated at a temperature typical of the freeze-out process--$M_1/25$ for definiteness. Each of them varies between $0$ and $1$ depending on how important the respective type of process is.  Thus,  $\zeta_{semi}^1\approx 1$  indicates that the $\phi_1$ relic density is mostly driven by semi-annihilations. Notice that they are normalized such that $\zeta_{anni}^1+\zeta_{semi}^1+\zeta_{conv}^1=1$. Analogous parameters can be defined for $\phi_2$. If a more detailed pictured is required of how the different  processes affect the relic density, {\tt micrOMEGAs} has the option, since its version 5.2,  to exclude from the calculation one or more  types of processes via the variable {\tt Excludefor2DM}.  We have used this option to perform several checks on our results.

Semi-annihilation processes will play a crucial role in our analysis so it is useful to get a feeling of how they compare against the usual annihilation processes. When $\phi_1$ annihilations are mediated by the typical Higgs portal, the thermally averaged cross section goes as 
\begin{align}
\sigma_v^{1100}&\sim \frac{\lambda_{S_1}^2}{16\pi M_1^2}\hspace{1cm}{\rm for}\,\,\, M_1\gg m_h.  
\end{align}
The semi-annihilation processes $\phi_1+\phi_1\to h+\phi_2$, on the other hand, feature a thermally averaged cross section 
\begin{align}
\label{eq:semisv}
\sigma_ v^{1120}&\sim \frac{\mu_{S1}^2v_H^2\lambda_{S_1}^2}{16\pi M_1^6}\hspace{1cm}{\rm for}\,\,\, \lambda_{S2}\ll \lambda_{S1},\,\,\, M_1\ll m_h.   
\end{align}
Since $\sigma_v^{1120}$ rapidly decreases with $M_1$, semi-annihilations are expected to stop being efficient at high values of  $M_1$. 

In the following section, we will impose the relic density constraint,
\begin{equation}
    \Omega_1+\Omega_2=\Omega_{DM},
\end{equation}
where $\Omega_{DM}$ is the observed  value of the dark matter density. The fraction of the total dark matter density that is accounted for by each dark matter particle is then given by the parameters
\begin{equation}
    \xi_i=\frac{\Omega_i}{\Omega_{DM}}\hspace{5mm}(i=1,2),
\end{equation}
with $\xi_1+\xi_2=1$. One of the main questions in two-component dark matter scenarios is determining what these  fractions are  (can they be comparable?), and they also affect the dark matter detection signals, as shown next.

\subsection{Direct and indirect detection}
The elastic scattering of the dark matter particles off nuclei are possible thanks to the Higgs portal interaction $\lambda_{Si}$,  just as  in the singlet scalar model~\cite{Silveira:1985rk,McDonald:1993ex,Burgess:2000yq}.  The expression for the spin-independent (SI) cross-section reads
\begin{align}
    \sigma_i^{{\rm SI}}&=\frac{\lambda_{Si}^2}{4\pi}\frac{\mu_R^2 m_p^2 f_p^2}{m_h^4 M_{i}^2}. 
\end{align}
where $\mu_R$ is the reduced mass, $m_p$ the proton mass and $f_p\approx 0.3$ is the quark content of the proton. But since we have two dark matter particles, the  quantity to be compared against the direct detection limits provided by the experimental collaborations is not the cross section itself but rather the product $\xi_i\sigma_i^{{\rm SI}}$. 

Such direct detection limits usually provide very strong constraints on Higgs-portal scenarios like the $Z_5$ model we are discussing.  For example,  in the limit $\Omega_2\ll \Omega_1$ and with the new $Z_5$ interactions switched off--where  the singlet complex scalar DM model~\cite{Silveira:1985rk,McDonald:1993ex,Burgess:2000yq} is recovered--we get that 
\begin{align}
\lambda_{S1}\sim 0.3 \left(\frac{M_1}{1\,{\rm TeV}}\right)\hspace{1cm}{\rm for}\,\,\, m_h\ll M_1,
\end{align}
in order to fulfill $\Omega_1=\Omega_{DM}$. Taking into account the upper limit set by the XENON1T collaboration \cite{Aprile:2018dbl} it follows that $M_1\gtrsim 2$ TeV (for a real scalar the lower limit is $\sim950$ GeV). 
Hence, for $M_1\lesssim2$ TeV the $Z_5$-invariant interactions must  be required in order to simultaneously satisfy the relic density constraint and current direct detection limits--a result we will numerically confirm in the next section. 
In our analysis, we will consider the current direct detection limit set by the XENON1T collaboration \cite{Aprile:2018dbl} as well as  the projected sensitivities of LZ~\cite{Akerib:2018lyp} and DARWIN \cite{Aalbers:2016jon}.

Regarding indirect detection, the relevant particle physics quantity switches from $\langle\sigma v\rangle$ to $\xi_i\xi_j\langle\sigma v\rangle_{ij}$, where  $\langle\sigma v\rangle_{ij}$ is the  cross section times velocity for the annihilation process of dark matter particles $i$ and $j$ into a certain final state. The main novelty in our model is the possible appearance of semi-annihilation processes involving two different dark matter particles, such as $\phi_1+\phi_1\to \phi_2+h$ or $\phi_1+\phi_2^\dagger\to \phi_1^\dagger+h$.
We will rely on the indirect detection limits and on the projected sensitivities reported by the Fermi collaboration from  observations of dShps  \cite{Ackermann:2015zua,Charles:2016pgz}.

\subsection{Parameter dependence}
To study how the different parameters affect the relic densities of the dark matter particles, we first define a \emph{reference} model in which most of these parameters are set to zero, and then switch them on, one by one, while comparing the resulting relic densities against the predictions of the reference model.  The non-zero parameters of the reference model are just four: the dark matter masses ($M_1$, $M_2$) and the Higgs-portal couplings ($\lambda_{S1}$, $\lambda_{S2}$). Note that even in this very simplified framework the relic densities are coupled via the Higgs-mediated processes $\phi_2+\phi_2^\dagger\leftrightarrow \phi_1+\phi_1^\dagger$ (see bottom-right panel in figure \ref{fig:conversion2}). 

For definiteness, in this section we set $\lambda_{S1}=\lambda_{S2}=0.1$, and examine two different values for the ratio $M_2/M_1$ (which can vary between $1$ and $2$): $1.2$ and $1.8$. In the following figures, the predictions of the reference model are shown in solid (green) lines. First, we are going to investigate the dependence of the relic densities on the dimensionless couplings ($\lambda_{31},\lambda_{32}, \lambda_{412}$) and then we move on to the dimensionful ones --$\mu_{S1}$ and $\mu_{S2}$.  

\subsubsection{The effect of $\lambda$'s}
\begin{figure}[t]
\includegraphics[scale=0.53]{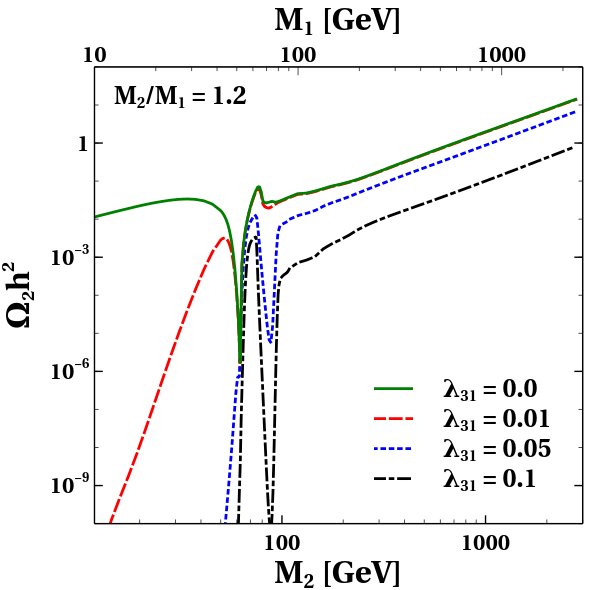}\hspace{0.2cm}
\includegraphics[scale=0.53]{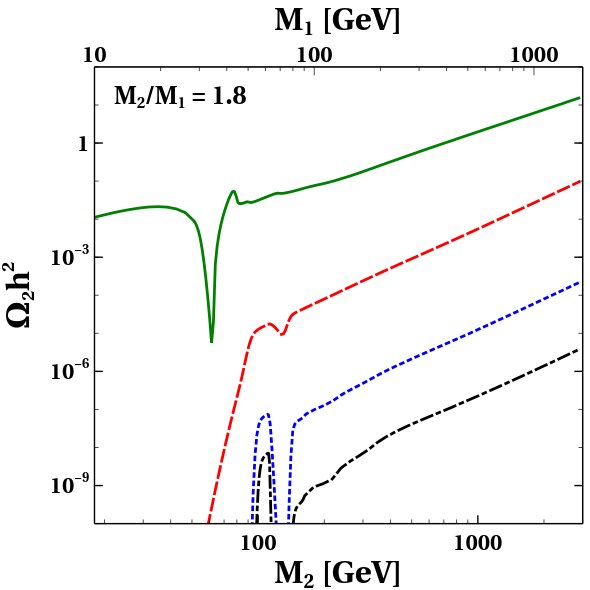}\\
\caption{The effect of $\lambda_{31}$ on $\Omega_2$ for two different values of $M_2/M_1$: $1.2$ (left panel) and $1.8$ (right panel).}
\label{fig:la31}
\end{figure}

 The dimensionless couplings--$\lambda_{31},\lambda_{32},\lambda_{412}$--induce the dark matter conversion processes shown in the top row of figure \ref{fig:conversion2}.  Neither semi-annihilations nor annihilations can be caused by these couplings. $\lambda_{31}$, for instance, leads to  the conversion processes $\phi_{1}+\phi_{2}\leftrightarrow \phi_1^\dagger+\phi_1^\dagger$ and their complex conjugates. During the $\phi_2$ freeze-out,  they contribute to the depletion of $\phi_2$ and should therefore reduce $\Omega_2$.  $\Omega_1$, on the other hand,  should hardly get modified unless $M_1\approx M_2$, when the kinematic suppression of $\phi_1+\phi_1\to \phi_1^\dagger+\phi_2^\dagger$ is alleviated. Figure \ref{fig:la31} shows $\Omega_2$ as a function of $M_2$ for $M_2/M_1=1.2~\mathrm{(left~panel)},1.8~\mathrm{(right~ panel)}$ and for different values of $\lambda_{31}$: $0.0$ (solid line), $0.01$ (dashed line), $0.05$ (dotted line), and $0.1$ (dash-dotted line). As expected, $\Omega_2$ decreases with $\lambda_{31}$ for both values of $M_2/M_1$. What is a bit surprising is the size of the effect. Notice, in fact, that even a value of $\lambda_{31}$ as small as $10^{-2}$ can modify $\Omega_2$ by several orders of magnitude. The reason behind this behavior is that the Boltzmann equation has a term of the  form ($Y_i=n_i/s$)
 \begin{align}
     \frac{dY_2}{dT}\propto \frac 12\sigma_v^{1211}Y_1Y_2
 \end{align}
 which exponentially suppresses the $\phi_2$ density over a range of temperatures. Thus, even a moderate value of $\sigma_v^{1211}$ can have a large impact on $\Omega_2$. And the larger $M_2/M_1$, the larger the suppression is. The other prominent feature in this figure is the dip observed above the Higgs resonance. It is actually caused by the usual bump in the $\phi_1$ relic density for $M_h/2\lesssim M_1\lesssim M_W$. Because the two relic densities are coupled, the increase in $Y_1$ provokes a reduction in $Y_2$. Notice, from the top axis, that the dip indeed occurs at the expected value of $M_1$.

\begin{figure}[tb]
\includegraphics[scale=0.53]{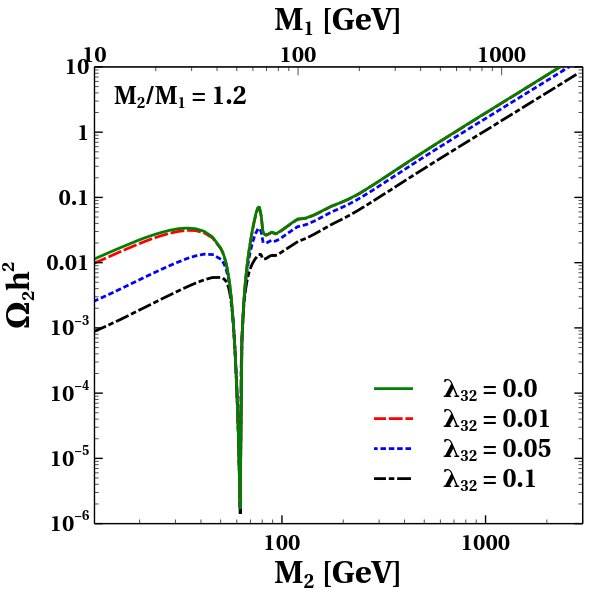}\hspace{0.2cm}
\includegraphics[scale=0.53]{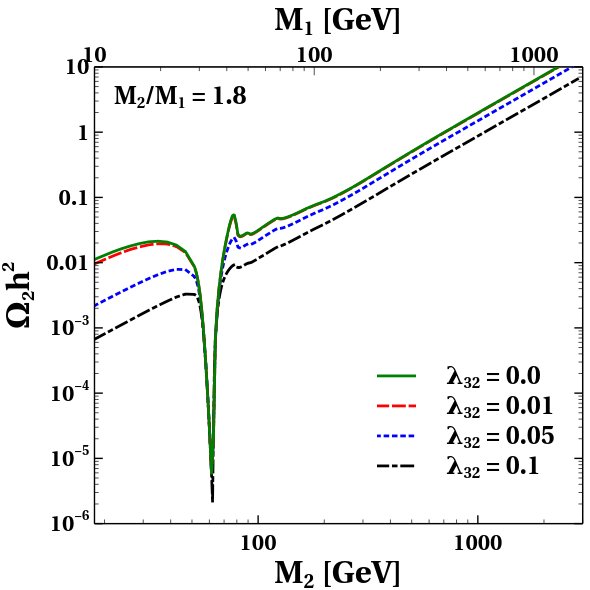}\\
\caption{The effect of $\lambda_{32}$ on $\Omega_2$ for two different values of $M_2/M_1$: $1.2$ (left panel) and $1.8$ (right panel).}
\label{fig:la32}
\end{figure}

Figure \ref{fig:la32} displays the effect  on $\Omega_2$ of $\lambda_{32}$. This coupling causes the conversion processes $\phi_{2}+\phi_{2}\leftrightarrow \phi_1+\phi_2^\dagger$, which should lead to a reduction of $\Omega_2$ while leaving $\Omega_1$ unaffected. From the figure, we see that $\Omega_2$ indeed decreases with $\lambda_{32}$ and that the effect is pretty much independent on $M_2/M_1$ --the two panels seem identical (they are not). For the values of $\lambda_{32}$ shown, the reduction in $\Omega_2$ reaches at most one order of magnitude.

\begin{figure}[tb]
\includegraphics[scale=0.53]{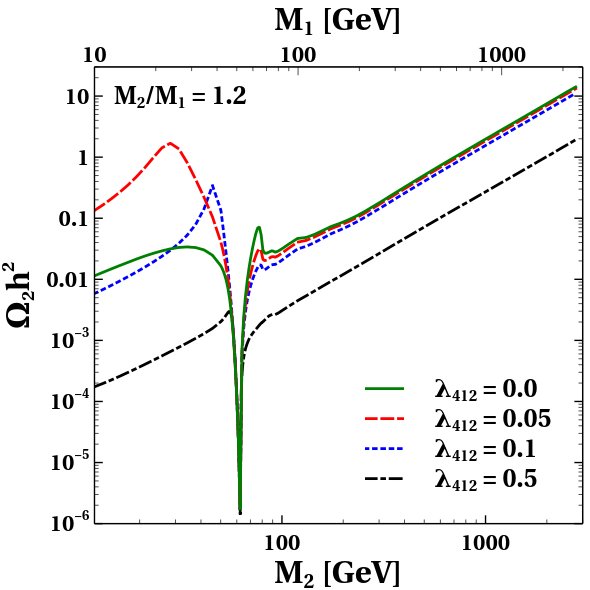}\hspace{0.2cm}
\includegraphics[scale=0.53]{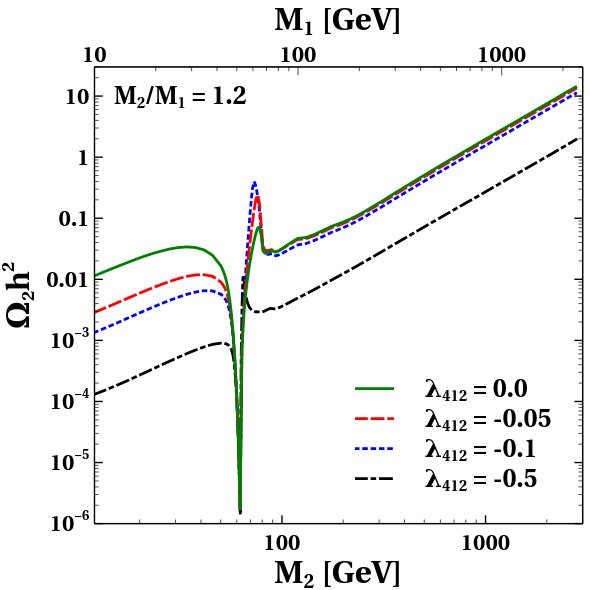}\\
\caption{The effect of $\lambda_{412}$ on $\Omega_2$ for $M_2/M_1=1.2$.  The difference between the left and the right panel is just the sign of $\lambda_{412}$.}
\label{fig:la412A}
\end{figure}

The last quartic coupling to be examined is $\lambda_{412}$, which should naively cause a reduction of $\Omega_2$ via the process $\phi_{2}+\phi_{2}^\dagger\rightarrow \phi_1+\phi_1^\dagger$.  Unlike the previous  processes, however, this one  receives an additional contribution from a Higgs-mediated diagram, and so interference effects between the two diagrams may play role and result in either an increase or a decrease of the relic density.  The Higgs-mediated amplitude is proportional to $\lambda_{S1}\lambda_{S2}$ and its sign changes (due to the propagator) at the Higgs resonance, $M_2\sim M_h/2$. Thus, the sign of $\lambda_{412}$ turns out to be relevant in the analysis. To illustrate these effects, figure \ref{fig:la412A} shows the relic density for $M_2/M_1=1.2$ and different values of $\lambda_{412}$ --they are positive in the left panel and negative in the right panel. From the figure the interference effects are evident. If $|\lambda_{412}|=0.05$, for instance, $\Omega_2$ is larger below the resonance and (sligthly) smaller above the resonance for a positive coupling (see left panel), but the other way around for a negative coupling (see right panel). On the other hand, if $\lambda_{412}$ is large enough, say $0.5$ (dash-dotted line), the interference effect is not as important (except very near the Higgs resonance) and the net result is that $\Omega_2$ decreases regardless of the sign of $\lambda_{412}$. For the couplings considered in figure \ref{fig:la412A}, the maximum variation in $\Omega_2$ amounts to  two orders of magnitude for masses below the Higgs resonance, and one order of magnitude above it. For $M_2/M_1=1.8$, the results are essentially identical, so they are not shown.

As we have seen, a common feature of the quartic interactions is that they \emph{mostly} affect the relic density of the heavier dark matter particle, $\Omega_2$. For the parameter values  we have considered in this section, the effect on  $\Omega_1$ is negligible.  
Thus, the  $\phi_1$ relic density is determined by the characteristic Higgs-mediated interactions of the singlet scalar model, and it is therefore expected to be subject to the same stringent direct detection constraints as that model. The trilinear couplings, $\mu_{S1}$ and $\mu_{S2}$, might influence $\Omega_1$ and help relax such constraints.  

\subsubsection{The effect of $\mu$'s}

\begin{figure}[tb]
\includegraphics[scale=0.53]{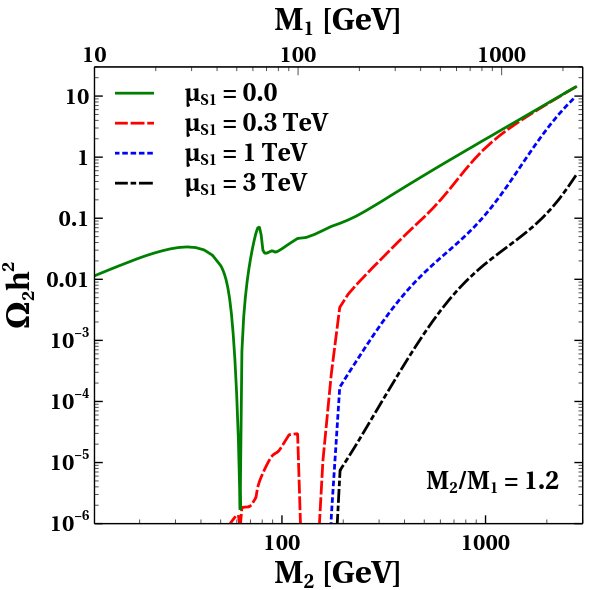}\hspace{0.2cm}
\includegraphics[scale=0.53]{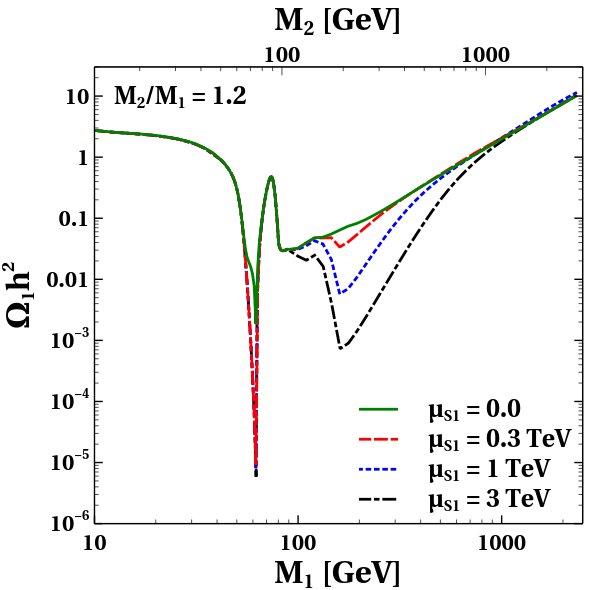}\\
\caption{The effect of $\mu_{S1}$ on $\Omega_2$ (left panel)  and  $\Omega_1$ (right panel) for $M_2/M_1=1.2$. }
\label{fig:mu1A}
\end{figure}

The trilinear couplings, $\mu_{S1}$ and $\mu_{S2}$, give rise to both  semi-annihilation and conversion processes --see  figures \ref{fig:semianni2} and \ref{fig:conversion2}. The semi-annihilation processes involve also one Higgs-dark matter coupling, either $\lambda_{S1}$ or $\lambda_{S2}$, and always feature a Higgs boson as an external particle. The conversion processes, on the other hand, depend only on $\mu_{Si}$ and are mediated by a dark matter particle in the t-channel. To illustrate how these processes alter the dark matter relic densities, in this section we  consider three possible values for $\mu_{Si}: 0.3, 1, 3~\mathrm{TeV}$.

\begin{figure}[tb]
\includegraphics[scale=0.53]{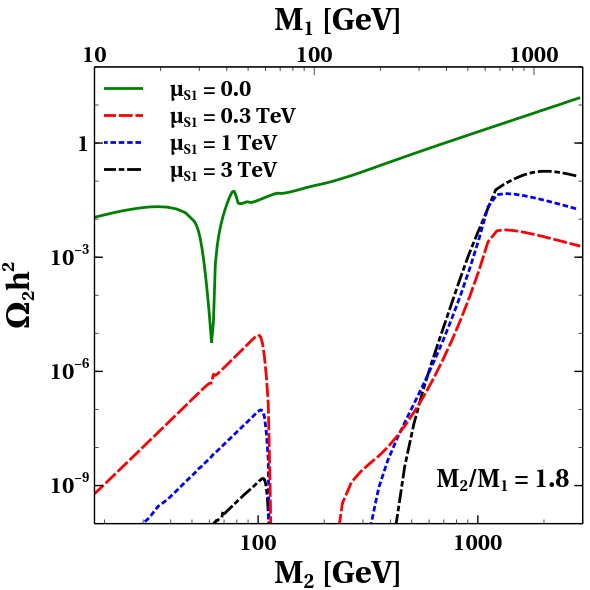}\hspace{0.2cm}
\includegraphics[scale=0.53]{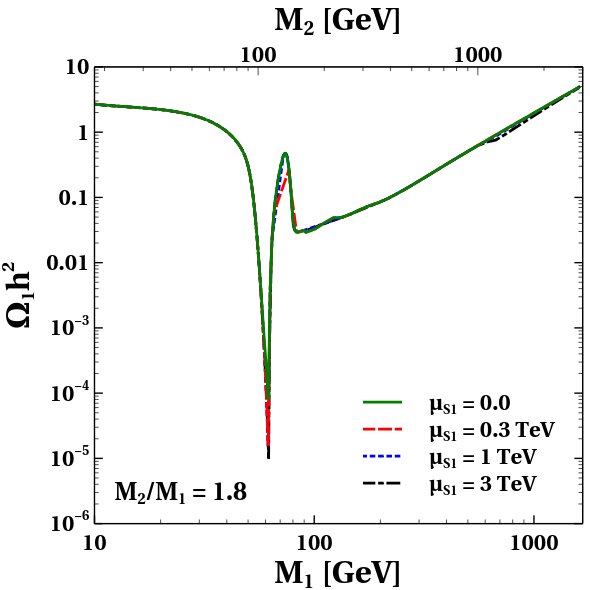}\\
\caption{The effect of $\mu_{S1}$ on $\Omega_2$ (left panel)  and  $\Omega_1$ (right panel) for $M_2/M_1=1.8$. }
\label{fig:mu1B}
\end{figure}

$\mu_{S1}$ induces the processes $\phi_1+\phi_2^\dagger\leftrightarrow \phi_1+h$ and $\phi_1+\phi_1\leftrightarrow \phi_2^\dagger+h$, the former affect only $\Omega_2$ while the latter may  affect both relic densities.  Figure \ref{fig:mu1A} displays $\Omega_i$ versus $M_i$ for different values of $\mu_{S1}$ and for $M_2/M_1=1.2$. From the left panel we see that $\Omega_2$ can be suppressed by orders of magnitude as a consequence of the exponential behaviour mentioned previously but now involving $\sigma_v^{1210}$. Notice also that $\Omega_2$ increases steeply as soon as the process $\phi_1+\phi_1\to \phi_2+h$ is kinematically open,  as observed in the figure. 
From the right panel, we notice instead that, at intermediate values of $M_1$, $\Omega_1$ can be reduced by up to two orders of magnitude. At low masses, the process $\phi_1+\phi_1\rightarrow \phi_2^\dagger+h$ is kinematically closed during $\phi_1$ freeze-out, so there is no effect on $\Omega_1$, in agreement with the figure.   
At high masses, it is instead the propagator that suppresses the $\phi_1+\phi_1\rightarrow \phi_2^\dagger+h$ diagram with respect to the standard Higgs-mediated processes. That is why there exists a finite range at moderate values of $M_1$ within which $\mu_{S1}$ can induce a reduction in $\Omega_1$ --see equation \ref{eq:semisv}. For $M_2/M_1=1.8$ (figure \ref{fig:mu1B}) the impact on $\Omega_1$ becomes negligible while $\Omega_2$ is even more suppressed. 

\begin{figure}[tb]
\includegraphics[scale=0.53]{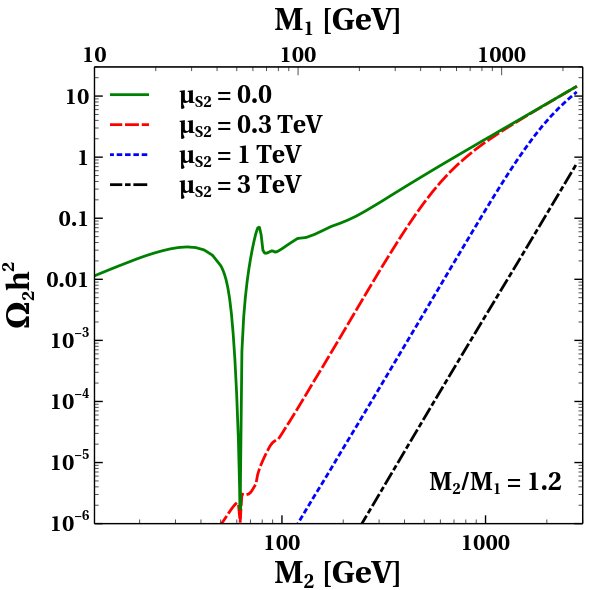}\hspace{0.2cm}
\includegraphics[scale=0.53]{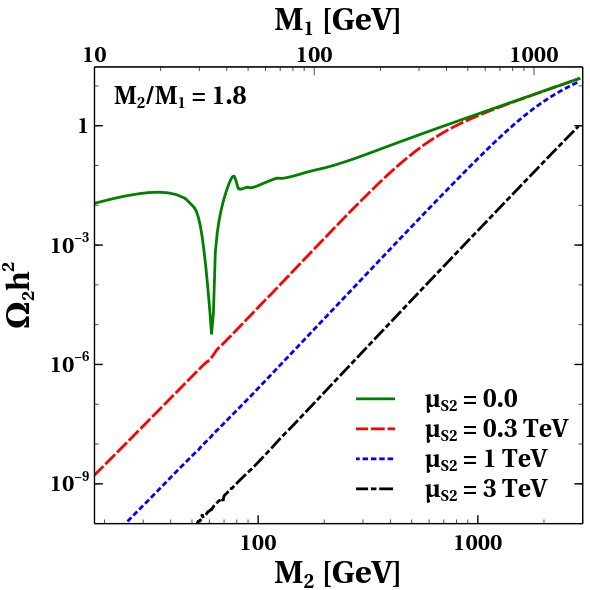}\\
\caption{The effect of $\mu_{S2}$ on $\Omega_2$ for $M_2/M_1=1.2$ (left panel) and $M_2/M_1=1.8$  (right panel). There is no appreciable effect on $\Omega_1$ for the values considered in this figure.}
\label{fig:mu2}
\end{figure}

Regarding the  $\mu_{S2}$-induced processes, they can affect $\Omega_2$ at low and intermediate masses as  shown in figure \ref{fig:mu2}. The only  process that may reduce the $\phi_1$ number density after $\phi_2$ freeze-out is $\phi_1+\phi_2\rightarrow \phi_2+h$ but it has a negligible effect on  $\Omega_1$ due to the small value of $\Omega_2$.

\section{The viable parameter space}
\label{sec:viable}
As we have seen, both relic densities may be modified by the new interactions allowed by the $Z_5$ symmetry. Now we want to explore in detail their implications on the viable parameter space of this model and on the dark matter detection prospects. To that end, we have randomly scanned the parameter space of the model and selected a large sample of points consistent with current data. In particular, they are compatible  with the limit on the invisible decays of the Higgs boson obtained from the LHC data \cite{Sirunyan:2018owy}, with the direct detection limits recently derived  by the XENON1T collaboration~\cite{Aprile:2018dbl} (we apply the corresponding recasted exclusion given by {\tt micrOMEGAs}~\cite{Belanger:2020gnr}) and with the dark matter density as measured by PLANCK~\cite{Aghanim:2018eyx}. While the PLANCK collaboration reports 
\begin{align}
    \Omega_{DM}h^2=0.1198\pm 0.0012,
\end{align}
the theoretical prediction of the relic density is not expected to be that precise. In our scans, we consider a model compatible with the above value if its relic density, as given by {\tt micrOMEGAs}, lies between $0.11$ and $0.13$, which amounts to about a $10\%$ uncertainty. In any case, our results are robust against plausible variations in such interval.

We have performed several random scans, varying just a subset of the free parameters of the model at a time so as to make the analysis simpler. In all the scans, the dark matter masses and the Higgs-portal couplings are varied in the following ranges:
\begin{align}
   &40\,{\rm GeV}\leq M_1 \leq 2\,{\rm TeV},\\
    &M_1< M_2 < 2M_1,\\
    &10^{-4}\leq |\lambda_{S1}|\leq 1,\\
    &10^{-3}\leq |\lambda_{S2}|\leq 1.
\end{align}
If these were the only parameters different from zero, the  viable points would all lie  at the Higgs resonance. The interplay between the relic density constraint and  the strong limits from direct detection searches would exclude the rest of the parameter space.   
And this conclusion still holds after  allowing $\lambda_{412}$ to be different from zero. Thus, it is up to the new $Z_5$ trilinear and quartic couplings to render  this model viable over most of the dark matter mass range. 

To bypass the direct detection bounds, the relic density of $\phi_1$ must be reduced by the new interactions. In the previous section, we saw that the parameter $\mu_{S1}$ can have this effect, so in our first scan  we set the dimensionless couplings as well as $\mu_{S2}$ to zero ($\lambda_{3i},\lambda_{412}=0, \mu_{S2}=0$) and vary $\mu_{S1}$ between $0.1$ TeV and $10$ TeV.  This upper limit on $\mu_{S1}$ is rather arbitrary but seems reasonable given that $M_1$ and $M_2$--the other dimensionful parameters of the model--take a maximum value of $2$ TeV and $4$ TeV respectively.

\begin{figure}[tb]
\centering
\includegraphics[scale=0.35]{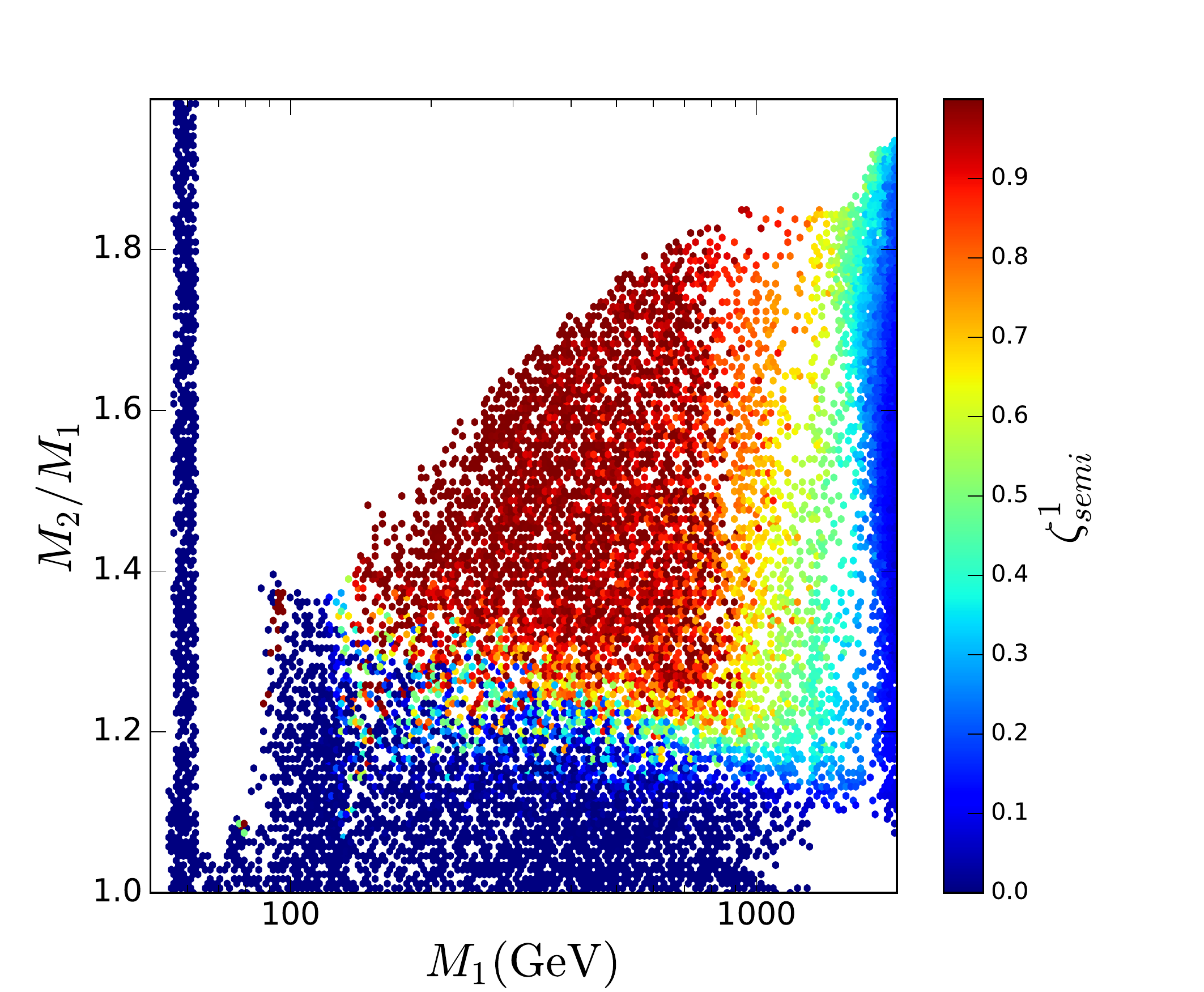}\hspace{0.2cm}
\includegraphics[scale=0.35]{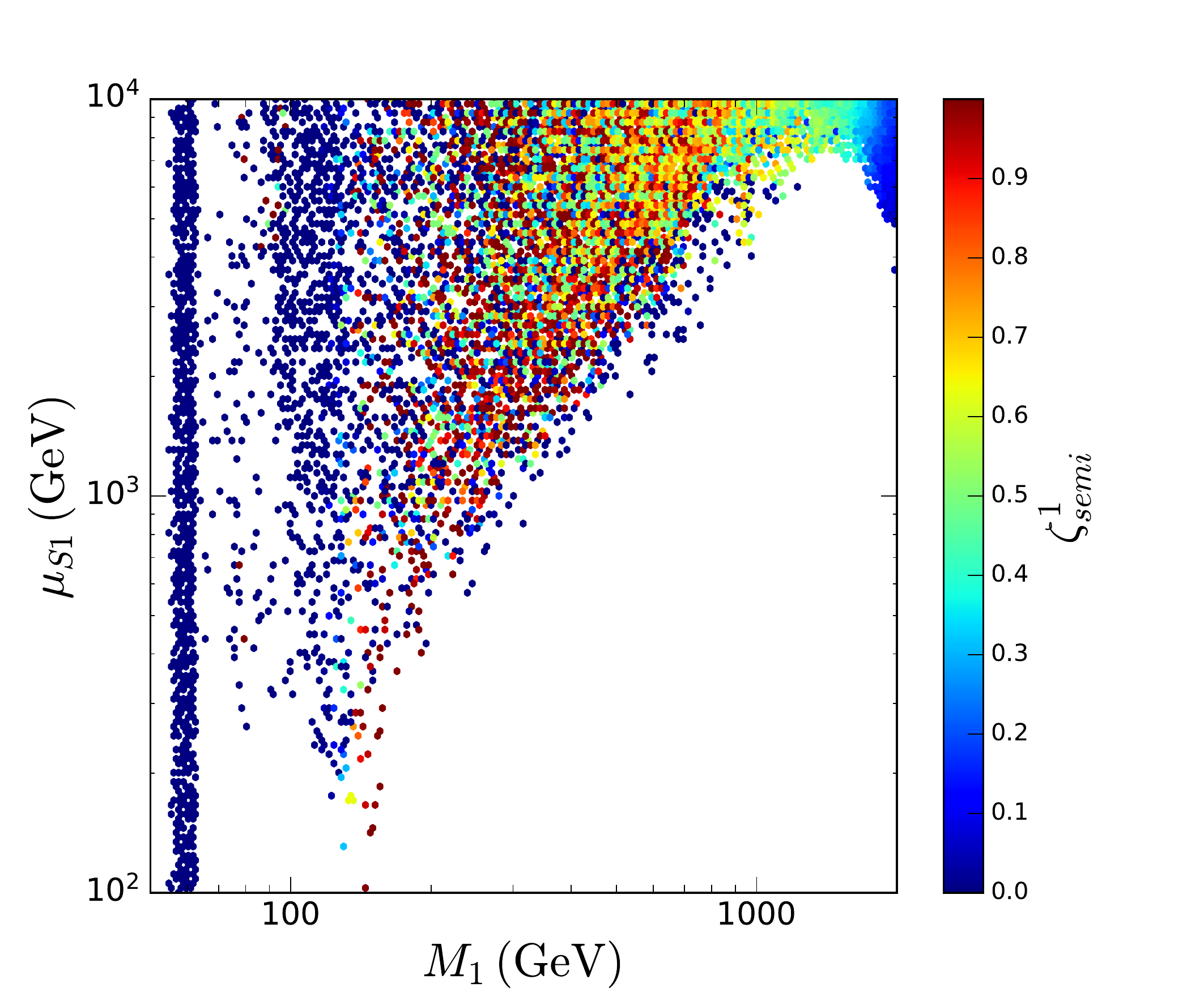}\\
\includegraphics[scale=0.35]{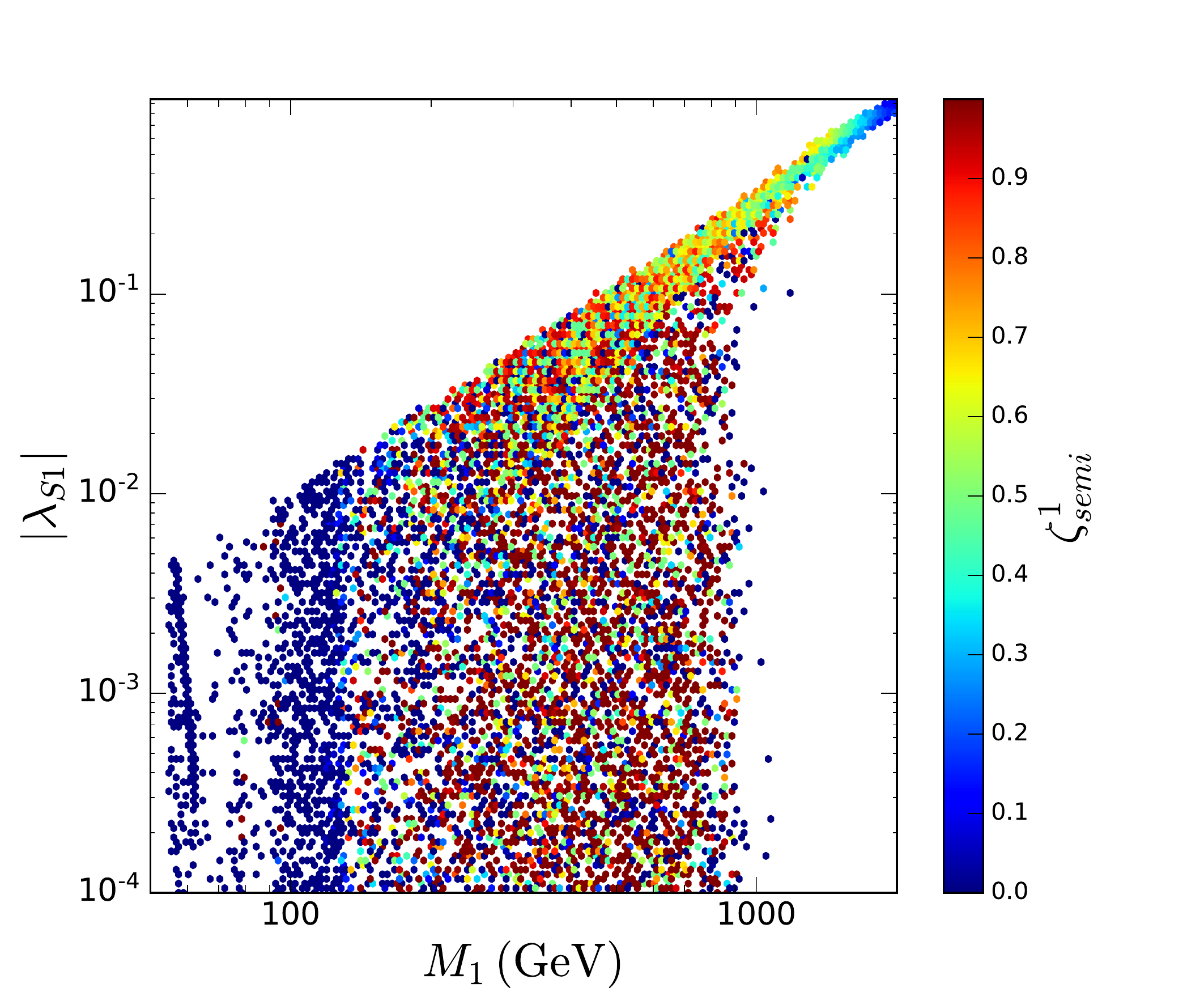}\hspace{0.2cm}
\includegraphics[scale=0.35]{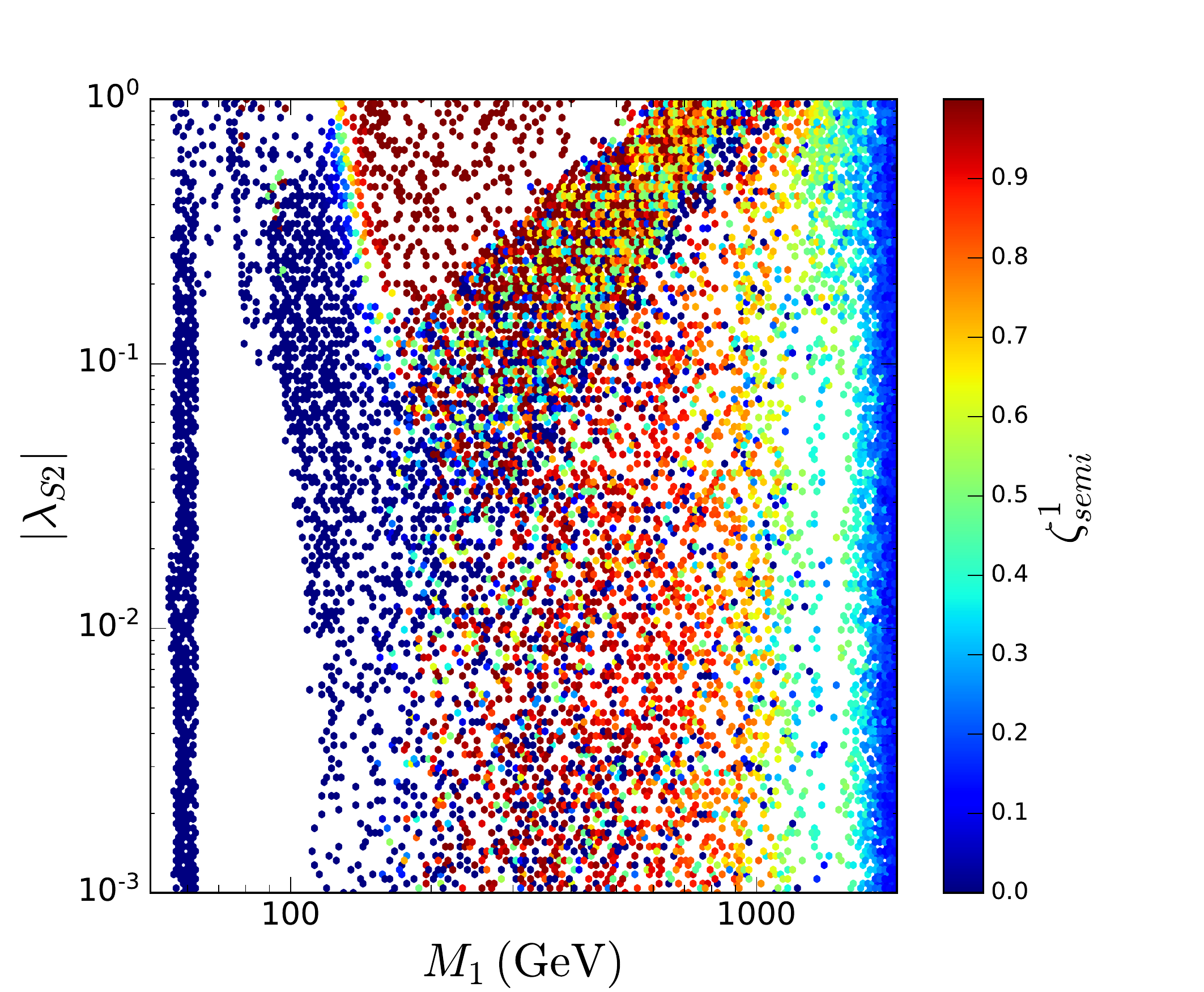}
\caption{Viable parameter space for $\mu_{S2}=0$ and $\lambda_{3i}=\lambda_{412}=0$.  The free parameters  ($M_2/M_1, \mu_{S2}, |\lambda_{Si}|$) are displayed as a function of $\phi_1$ mass and characterized by the semi-annihilation fraction $\zeta^1_{semi}$.
}
\label{fig:scanmu1A}
\end{figure}

The resulting viable parameter space  is shown in figure \ref{fig:scanmu1A}.  Notice that the viable points cover the entire spectrum of dark matter masses, from the Higgs resonance up to the maximum value considered in the scan. This is one of our main results. From the top-left panel, we see that the ratio $M_2/M_1$ varies over a wide range, indicating that the dark matter particles do not need to be degenerate.   In these plots, the relevance of semi-annihilation processes is color-coded in terms of $\zeta_{semi}^1$ --see equation \ref{eq:semi}.
Semi-annihilations are essential in the intermediate mass region ($200<M_1/\mathrm{GeV}<1000$), with most points featuring $\zeta_{semi}^1>0.75$. At low masses ($M_1\lesssim 200$ GeV), semi-annihilations are kinematically suppressed whereas at high masses ($M_1\gtrsim 1.5$ TeV) they are required but not as efficient. In fact, the minimum value of $\mu_{S1}$ increases with $M_1$ up to about $1$ TeV (top-right panel), when it reaches the maximum value allowed in the scan ($10$ TeV). Had we considered higher values of $\mu_{S1}$, semi-annihilations would  have remained significant to  larger dark matter masses. The Higgs-portal couplings are shown in the bottom panels. $|\lambda_{S1}|$ can vary over orders of magnitude while semi-annihilations are relevant, $M_1< 1$ TeV, but from then on annihilations become important and $|\lambda_{S1}|$ is therefore restricted to a narrow band, reaching $1$ for $M_1\sim 2$ TeV.  
The distribution of $|\lambda_{S2}|$ tends to be concentrated toward higher values (see bottom-left panel), with a significant fraction of models featuring $|\lambda_{S2}|\geq 0.1$ for $M_1<1\,\mathrm{TeV}$ ($M_2<2\,\mathrm{TeV}$). As we will see, this result has important implications for the dark matter detection prospects in this model.

\begin{figure}[tb]
\centering
\includegraphics[scale=0.52]{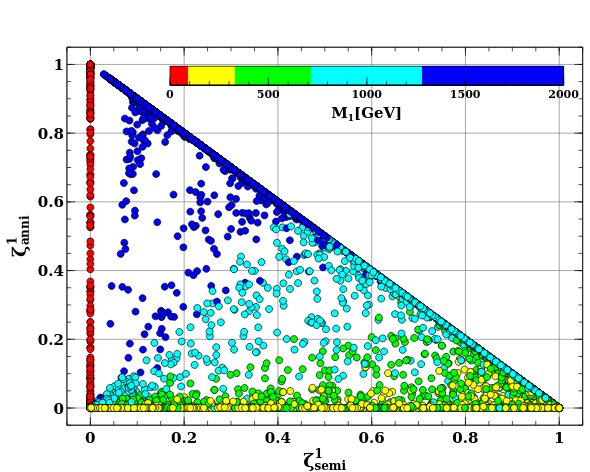}\hspace{5mm}
\includegraphics[scale=0.35]{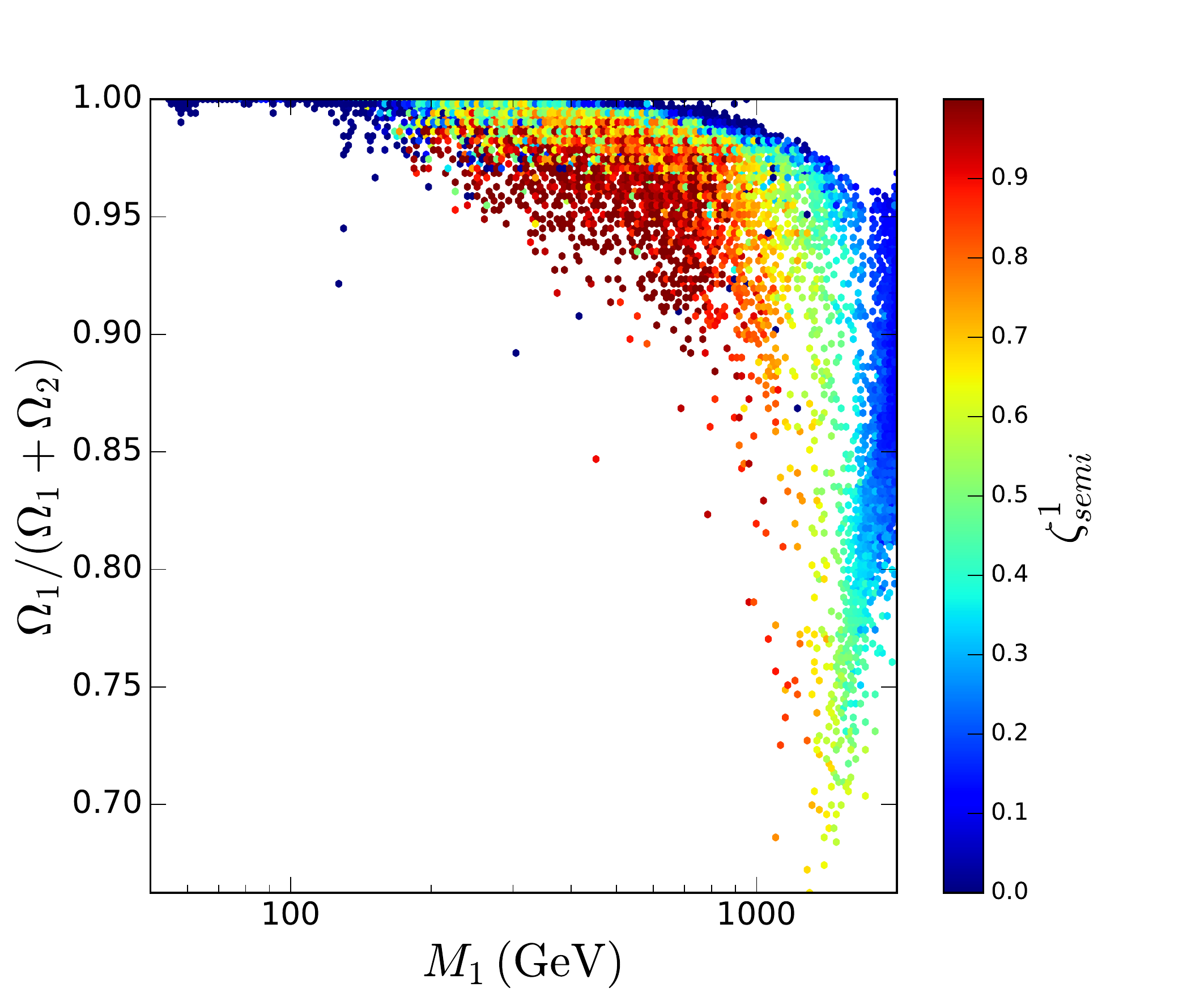}
\caption{Semi-annihilation and annihilation fractions (left panel) and the relative contribution of $\phi_1$ to the total DM relic abundance as a function of $M_1$ (right panel) for the scan with  $0.1\leq\mu_{S1}\leq10$ TeV, $\mu_{S2}=0$ and $\lambda_{3i}=\lambda_{412}=0$.
}
\label{fig:scanmus1Omega}
\end{figure}

We already learned, from figure \ref{fig:scanmu1A}, that semi-anninilations are important in the intermediate mass region. But what about conversions and annihilations? The left panel of figure \ref{fig:scanmus1Omega} shows the viable models in the plane $(\zeta_{semi}^1,\zeta_{anni}^1)$--see equation \ref{eq:semi}--with the color indicating the value of $M_1$.  The value of $\zeta_{conv}^1$ can be deduced from the figure by noting that $\zeta_{semi}^1+\zeta_{anni}^1+\zeta_{conv}^1=1$. By definition, all models have to lie either  inside the triangle with vertices $(0,0)$, $(1,0)$ and $(0,1)$, when all three types of processes contribute to the relic density; or along its edges, when the contribution from one type is negligible. This latter case is seen to be the most common, with the negligible type depending on $M_1$: semi-annihilations at low masses, annihilations at intermediate values, and conversions at high masses.

\begin{figure}[tb]
\centering
\includegraphics[scale=0.35]{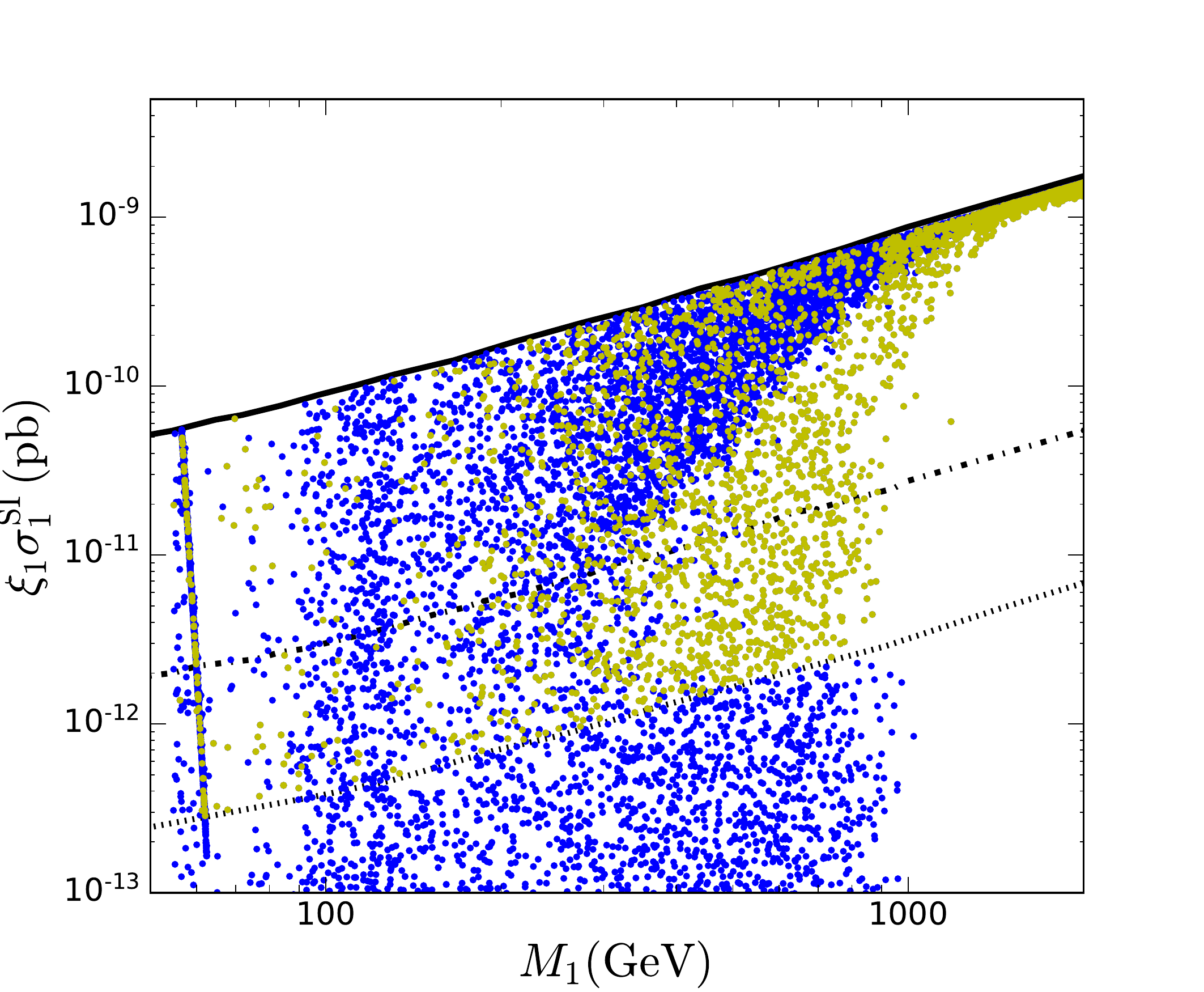}\hspace{0.2cm}
\includegraphics[scale=0.35]{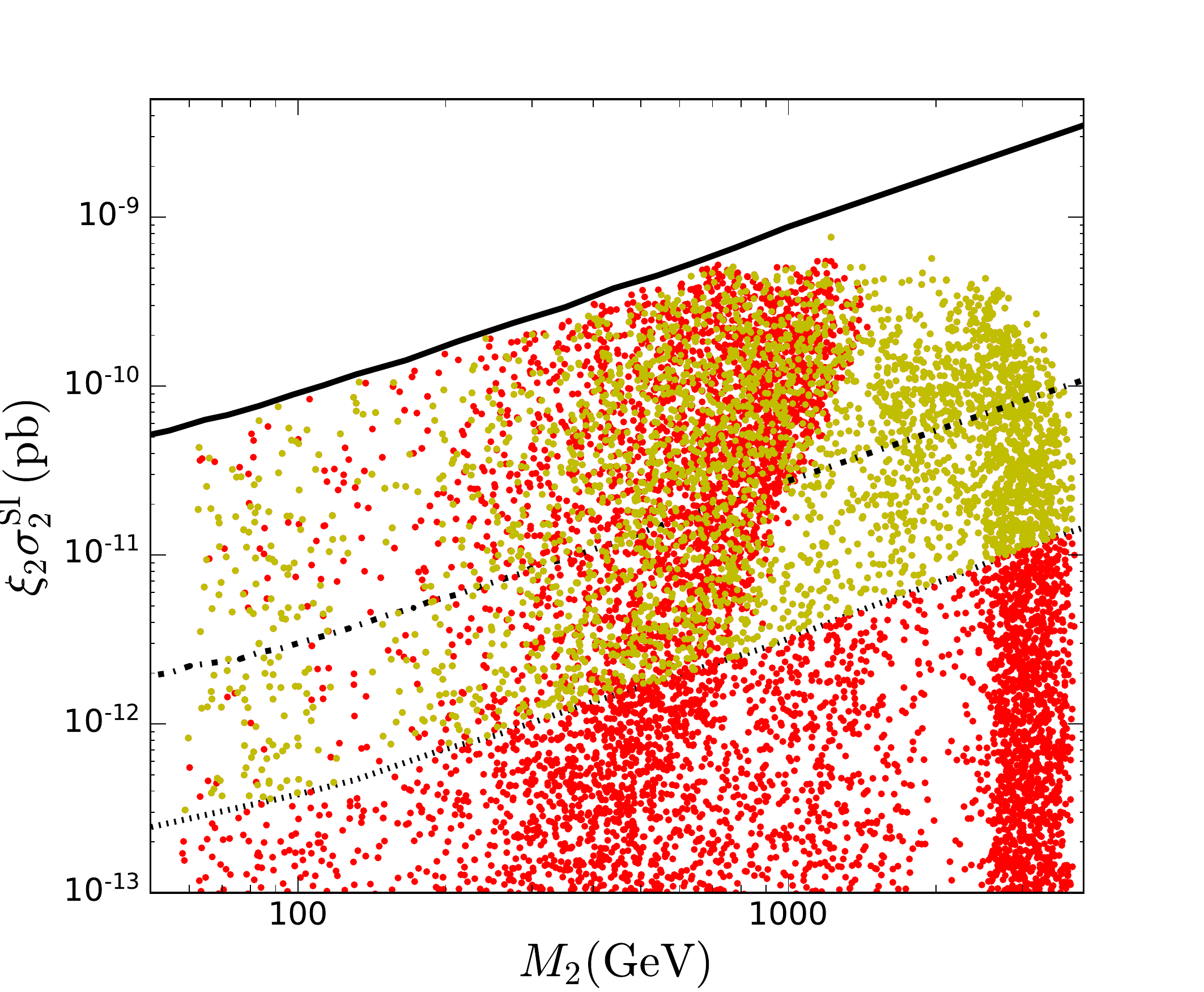}\hspace{0.2cm}
\caption{Spin-independent cross-sections for elastic scattering of $\phi_i$ with nuclei scaled by $\xi_i$ in the scan with $\mu_{S1}\neq0$. The solid line is the upper limit set by XENON1T collaboration \cite{Aprile:2018dbl} while the dot-dashed and dotted lines show the projected sensitivity of LZ~\cite{Akerib:2018lyp} and DARWIN \cite{Aalbers:2016jon} experiments. Yellow points indicate that both DM particles lay within the sensitivity region of DARWIN.  
}
\label{fig:scanmu1DD}
\end{figure}

Regarding the contributions of the two dark matter particles to the total density, we see, from the right panel of  figure \ref{fig:scanmus1Omega}, that $\phi_1$ always gives the dominant contribution. It accounts for more than $70\%$ of the dark matter density and in most points for more than $95\%$ of it. In numerous cases $\Omega_2$ turns out to be several orders of magnitude smaller than $\Omega_1$.  The fact that the lighter dark matter particle usually accounts for the bulk of the dark matter density is one of our most important results. 

At first sight, this distribution of the dark matter densities may seem to imply that the $Z_5$ model \emph{effectively} becomes, at present, a one-component dark matter model --that $\phi_2$, having a small density, can be ignored. But this is not so. From figure \ref{fig:scanmu1DD} we  see that either dark matter particle may be observed in future direct detection experiments. The solid line shows the current limit from XENON1T while the dashed and dotted lines correspond to the expected sensitivities of   LZ~\cite{Akerib:2018lyp} and DARWIN \cite{Aalbers:2016jon} respectively.  What is happening with $\phi_2$ is that its smaller density can be compensated by its larger coupling to the Higgs (see figure \ref{fig:scanmu1A}), resulting in a sizable signal. The feasibility of detecting a subdominant component of the dark matter has been noted before \cite{Duda:2001ae,Duda:2002hf}, but it seems to have been largely forgotten. In the $Z_5$ model, this possibility arises naturally.

For $\phi_1$, two regions can be clearly distinguished (see the left panel). If $M_1\gtrsim 1$ TeV --when the semi-annihilations are not as efficient-- all viable points are at the brink of being detected, lying just below the current XENON1T limit. If $M_1\lesssim 1$ TeV instead, the (scaled) elastic scattering cross section varies over orders of magnitude, with some points close to the current limit and others located below the expected sensitivity of future experiments. For $\phi_2$ (right panel),  most of the detectable points feature $M_2\lesssim 1.5$ TeV while the non-detectable models are often characterized by a small value of $\xi_2=\Omega_2/\Omega_{DM}$. In this figure, the yellow points denote the viable models for which \emph{both} dark matter particles are expected to yield signals in future direct detection experiments. If observed, such signals would rule out the  \emph{one dark matter particle paradigm}  and open the way for multi-component dark matter scenarios such as the $Z_5$ model we are discussing.

\begin{figure}[tb]
\includegraphics[scale=0.5]{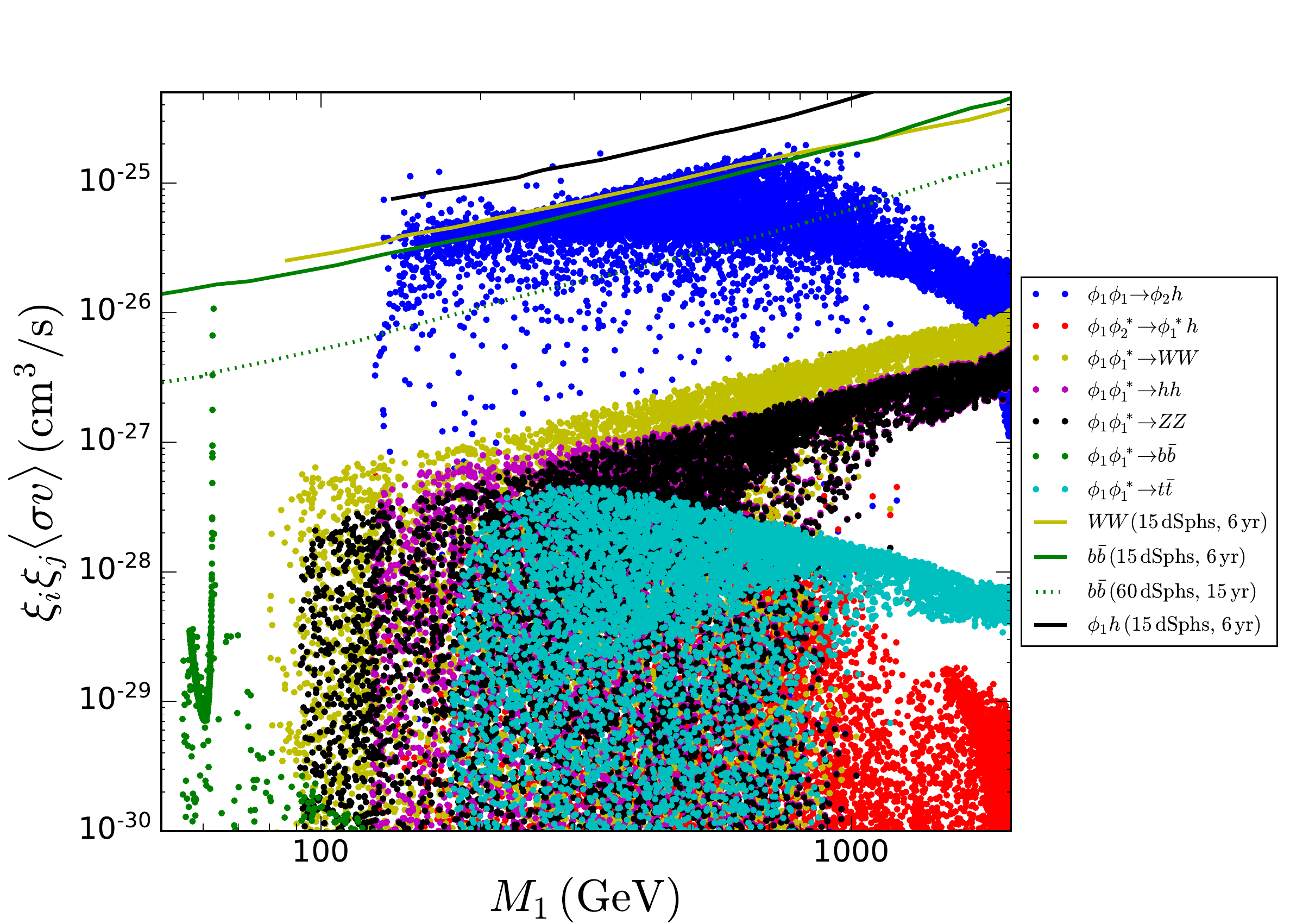}
\caption{Dark matter annihilation rates for the viable models in the scan with $\mu_{S1}\neq 0$. The solid-green (solid-yellow) line shows current limit of $\phi_1$ self-annihilation into $b\bar{b}$ ($W^+W^-$) reported by the Fermi collaboration from 6 years of observation and 15 dwarf spheroidal galaxies (dSphs) \cite{Ackermann:2015zua}, while the dotted-green line represents the projected sensitivity for 45 dSphs and 15 years of observation \cite{Charles:2016pgz} which serves as an estimate of the corresponding $W^+W^-$ sensitivity since both bounds for 6 years as 15 dSphs are similar at high DM masses. Moreover, for comparison purposes the upper limit on the semi-annihilation process $\phi_1\phi_1^{*} \to \phi_1 h$ ~\cite{Queiroz:2019acr} is also displayed.  
}
\label{fig:scanmu1ID}
\end{figure}

With respect to indirect detection, the most relevant dark matter annihilation channels are displayed in figure \ref{fig:scanmu1ID} with their respective scaled cross sections.  For comparison, the current limits \cite{Ackermann:2015zua} for certain final states are also shown (solid lines) as well as the projected sensitivity \cite{Charles:2016pgz} for $b\bar b$ (dotted line). The semi-annihilation process $\phi_1+\phi_1\to \phi_2+h$ turns out to be the most relevant one, with a cross section that can reach $10^{-25}\mathrm{cm^3/s}$. The experimental limit on such a process will depend also on $M_2$ and has not been derived in the literature. A related process which has been considered is $\phi_1+\phi_1\to \phi_1+h$, whose limit is shown in the figure as a solid black line \cite{Queiroz:2019acr}. Since $M_2>M_1$, the limit on $\phi_1+\phi_1\to \phi_2+h$ should be weaker. Due to  the $\xi_2$ suppression and its higher mass, the indirect detection signals involving $\phi_2$ are less promising. Indirect detection experiments, therefore, do not constrain the viable parameter space of this model.

\begin{figure}[h!]
\centering
\includegraphics[scale=0.35]{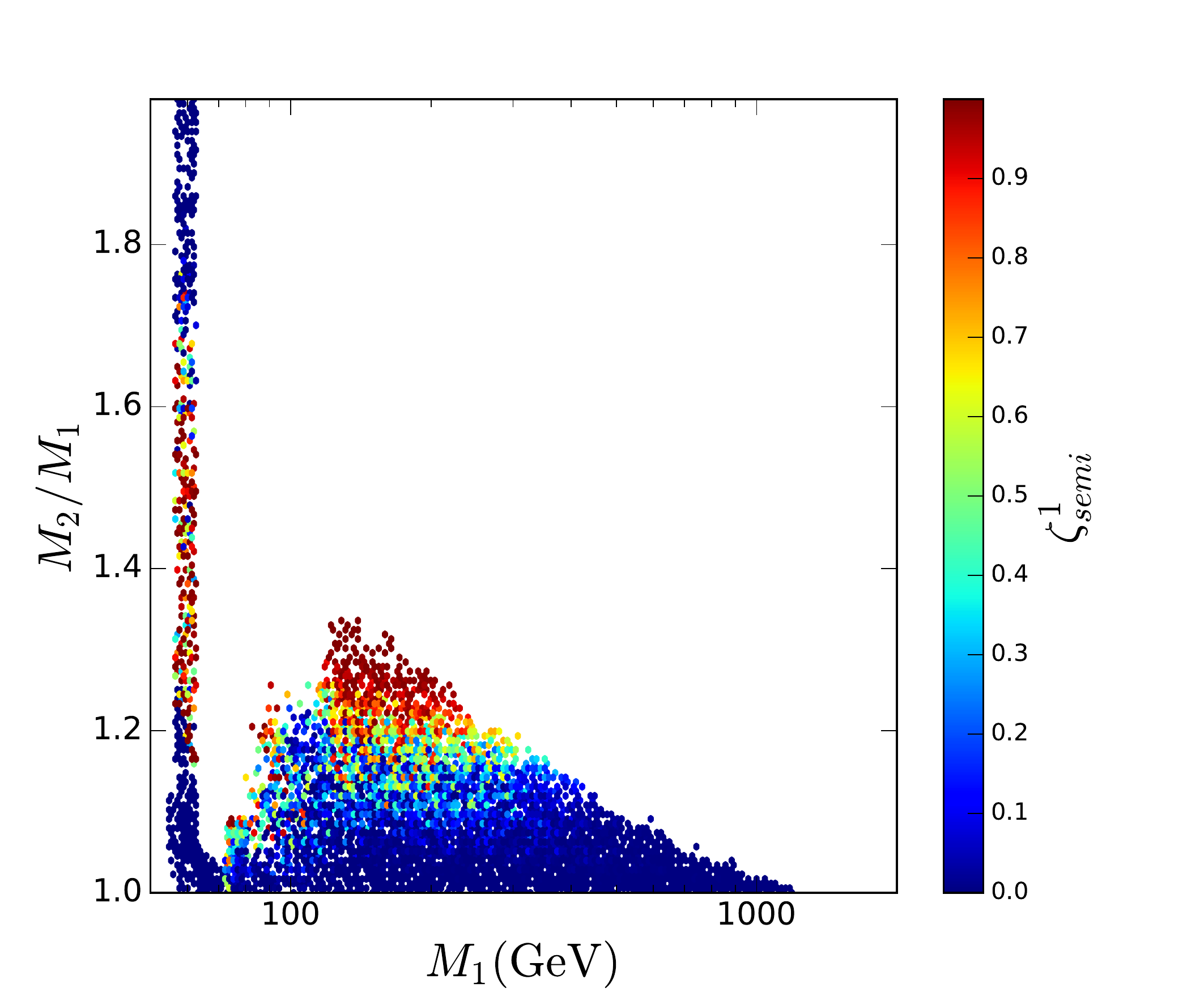}\hspace{0.2cm}
\includegraphics[scale=0.35]{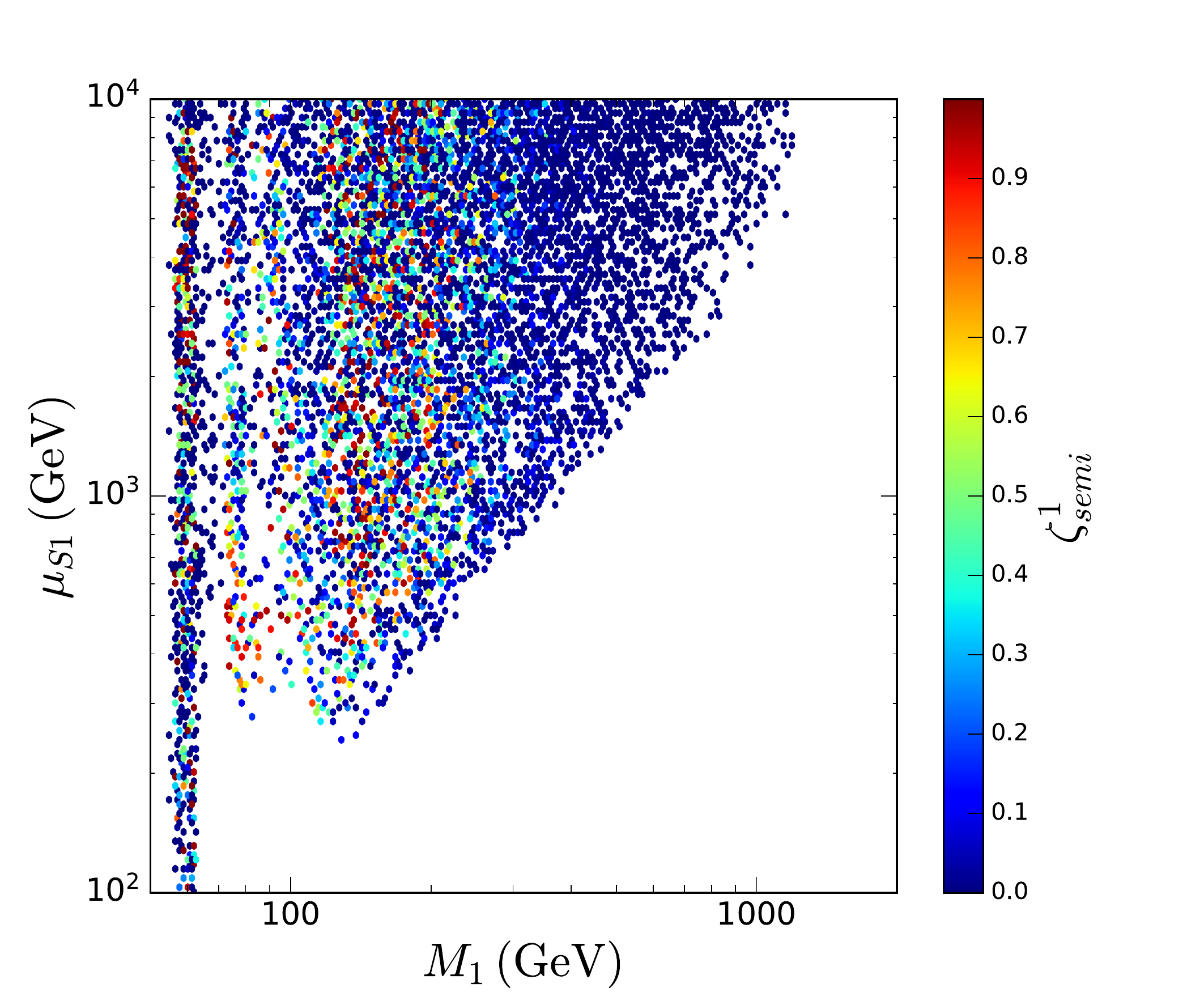}\\
\includegraphics[scale=0.45]{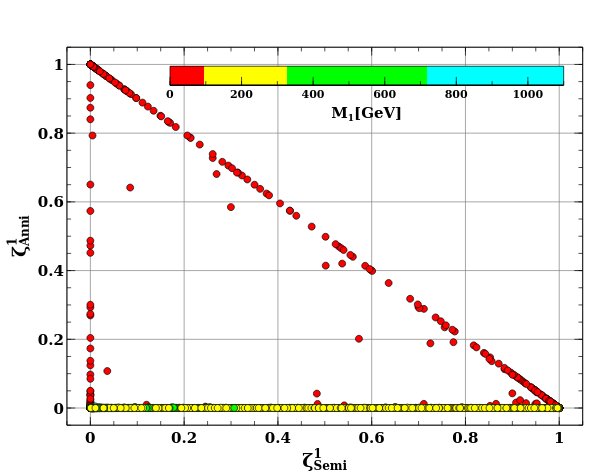}\hspace{5mm}
\includegraphics[scale=0.32]{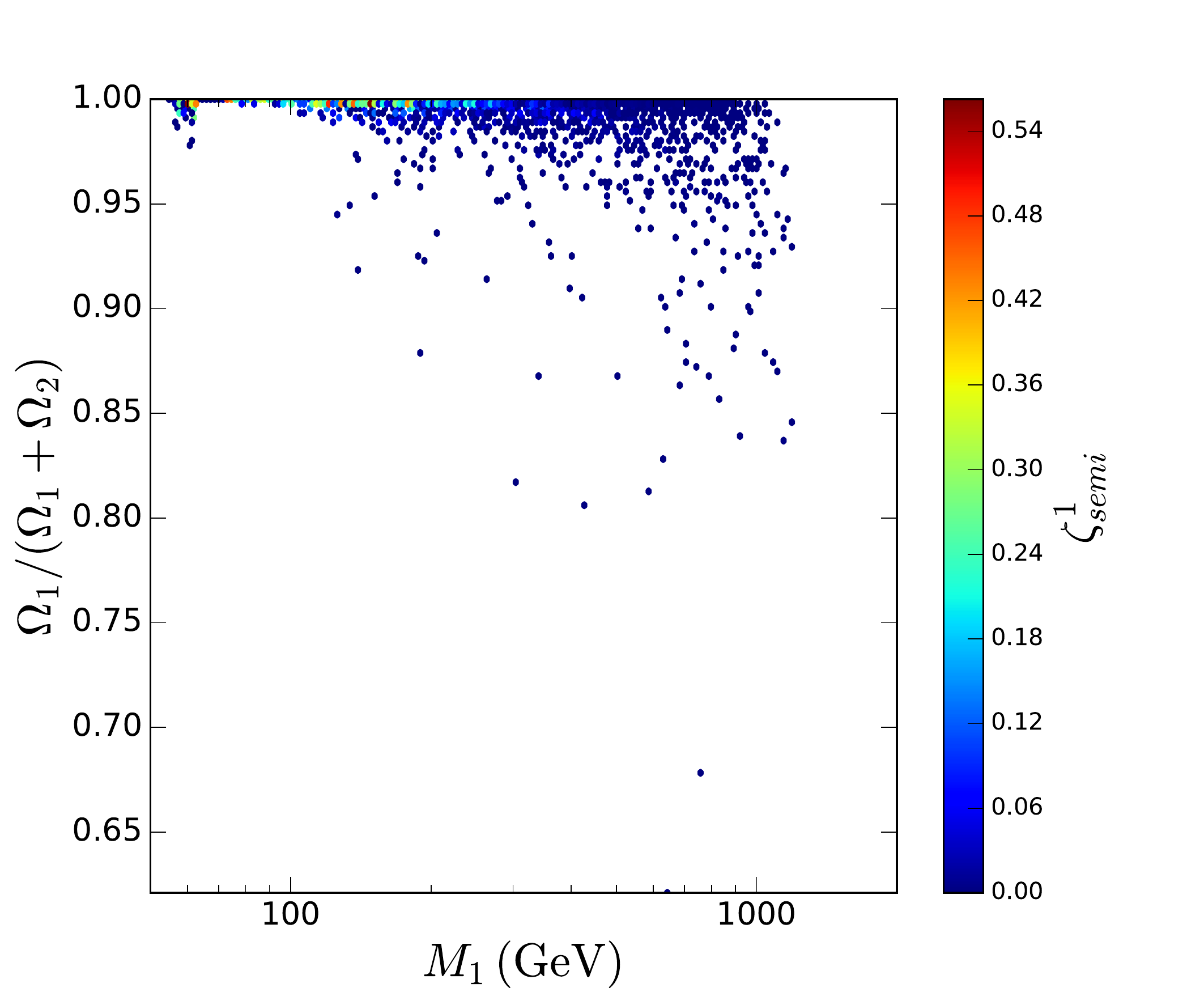}\\
\includegraphics[scale=0.35]{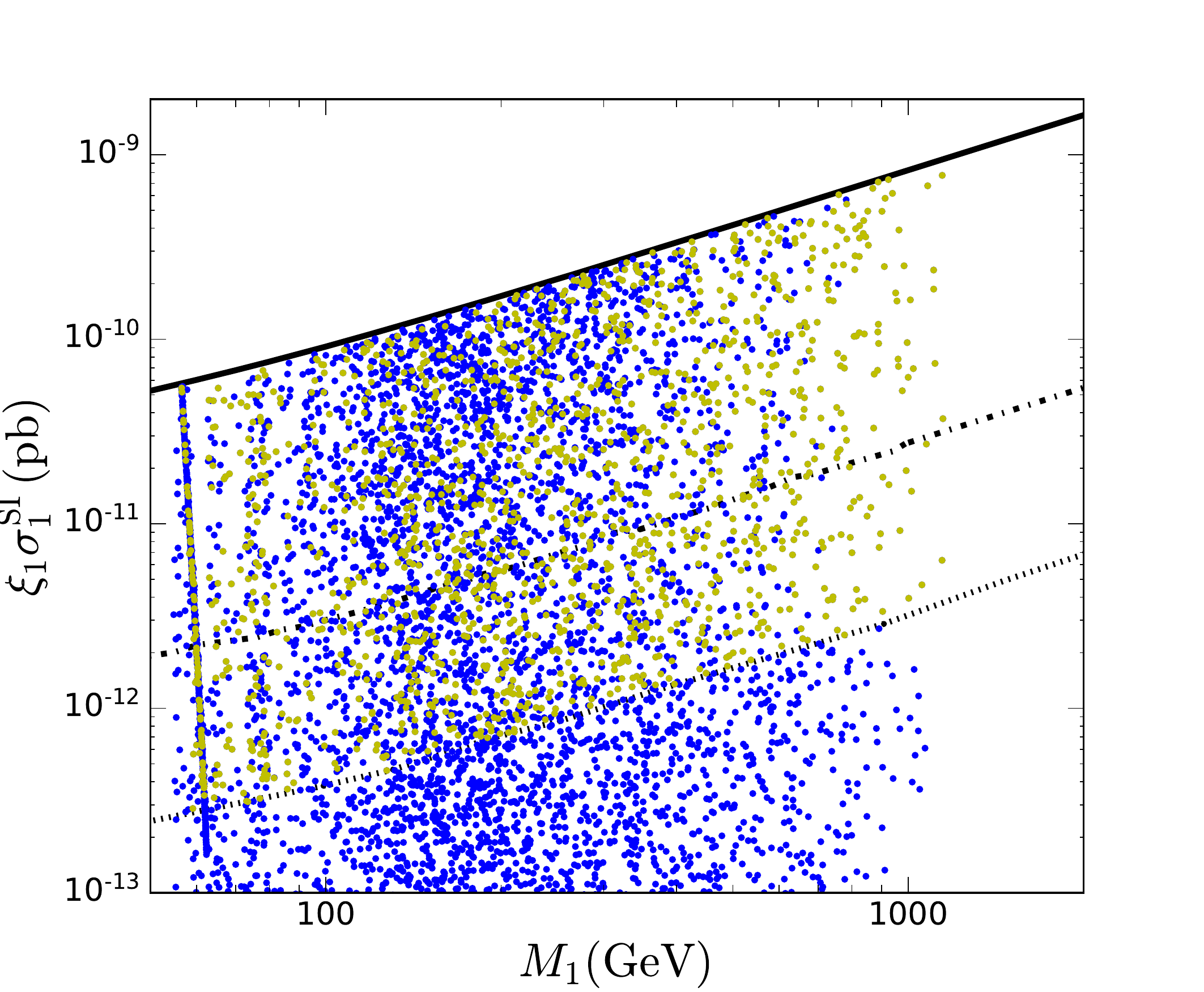}\hspace{0.2cm}
\includegraphics[scale=0.35]{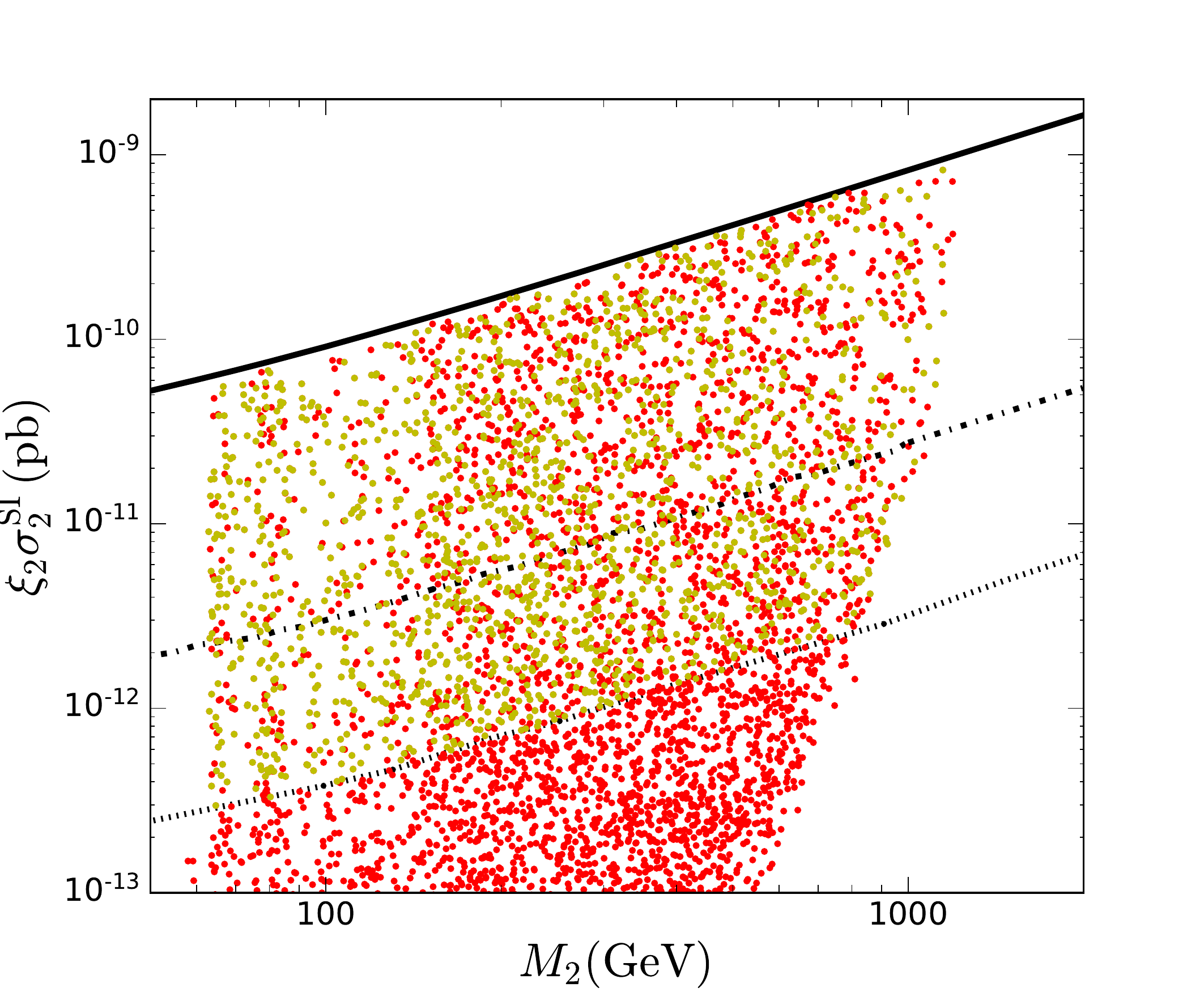}
\caption{Results for the scan with $0.1\leq\mu_{S2}\leq10$ TeV, $\mu_{S1}=0$ and $\lambda_{3i}=\lambda_{412}=0$. Top panels:  the viable parameter space; center panels: annihilation fraction vs semi-annihilation fraction and relative contribution of $\Omega_1$ to $\Omega_{DM}$; bottom panels: SI cross-sections scaled by $\xi_i$ where the solid line is the upper limit set by XENON1T collaboration \cite{Aprile:2018dbl}  and the dot-dashed (dotted) line is the projected sensitivity of LZ~\cite{Akerib:2018lyp} (DARWIN \cite{Aalbers:2016jon}) experiment.  
}
\label{fig:scanmuS2}
\end{figure}

Let us summarize what we have found with the scan for  $\mu_{S1}\neq 0$: $i)$ the model becomes viable over the entire range of dark matter masses, $M_1<2~\mathrm{TeV}$; $ii)$ $\phi_1$, the lighter dark matter particle, accounts for most of the dark matter density; $iii)$ direct detection experiments offer great prospects to test this model, including the possibility of observing signals from \emph{both} dark matter particles. As we wil see, $ii)$ and $iii)$ are actually generic features of the viable parameter space of the $Z_5$ model.

So far, we have examined the effect of $\mu_{S1}$ on the viable parameter space of the model, but  what about the other couplings? Even if their effect on $\Omega_1$ could not be observed
in the examples given in the previous section,  they may be present under certain circumstances. For that reason, we also did scans varying $\mu_{S2}$ and the dimensionless couplings.

The results for the scan with $\mu_{S2}\neq 0$ are shown in figure \ref{fig:scanmuS2}. In this case, we set the dimensionless couplings as well as $\mu_{S1}$ to zero ($\lambda_{3i},\lambda_{412}=0, \mu_{S1}=0$) and vary $\mu_{S2}$ between $0.1$ TeV and $10$ TeV. Three crucial differences are observed with respect to the results from the $\mu_{S1}$ scan. First, there is a range of dark matter masses, above 1.1 TeV approximately, for which no viable models are found (top panels). Second, the dark matter masses have to be degenerate, with $M_2/M_1$ reaching a maximum value of about $1.3$ for $M_1\sim 100~\mathrm{GeV}$ and decreasing steeply with $M_1$ (top-left panel). Finally,  it is the conversion process $\phi_1+\phi_1\to \phi_2+\phi_2$--mediated by a $\phi_2$--that reduces the $\phi_1$ relic density over most of the viable range of $M_1$, with semi-annihilations being relevant only at low masses (top  and center-left panels). 

But there are also important similarities with the previous scan. The dark matter density is still dominated by the lighter component ($\phi_1$) for all viable points (center-right panel), and direct detection experiments remain the most promising way to test this scenario in the near future (bottom panels). In particular, a significant fraction of models predict detectable signals from both dark matter particles (yellow points). Discriminating such signals would, however, become more challenging in this case due to the degeneracy between the dark matter particles.

\begin{figure}[tb]
\centering
\includegraphics[scale=0.35]{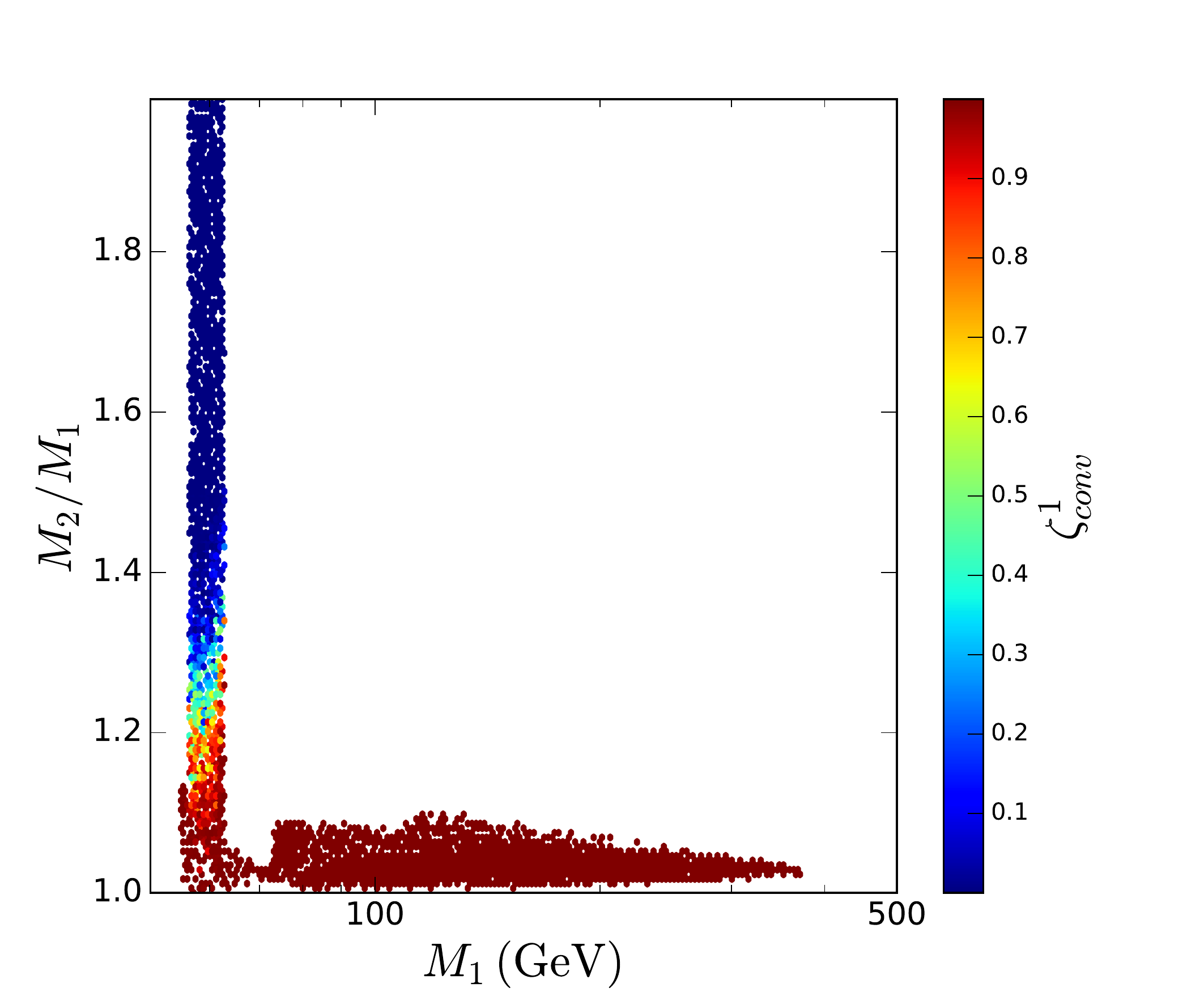}\hspace{0.2cm}
\includegraphics[scale=0.35]{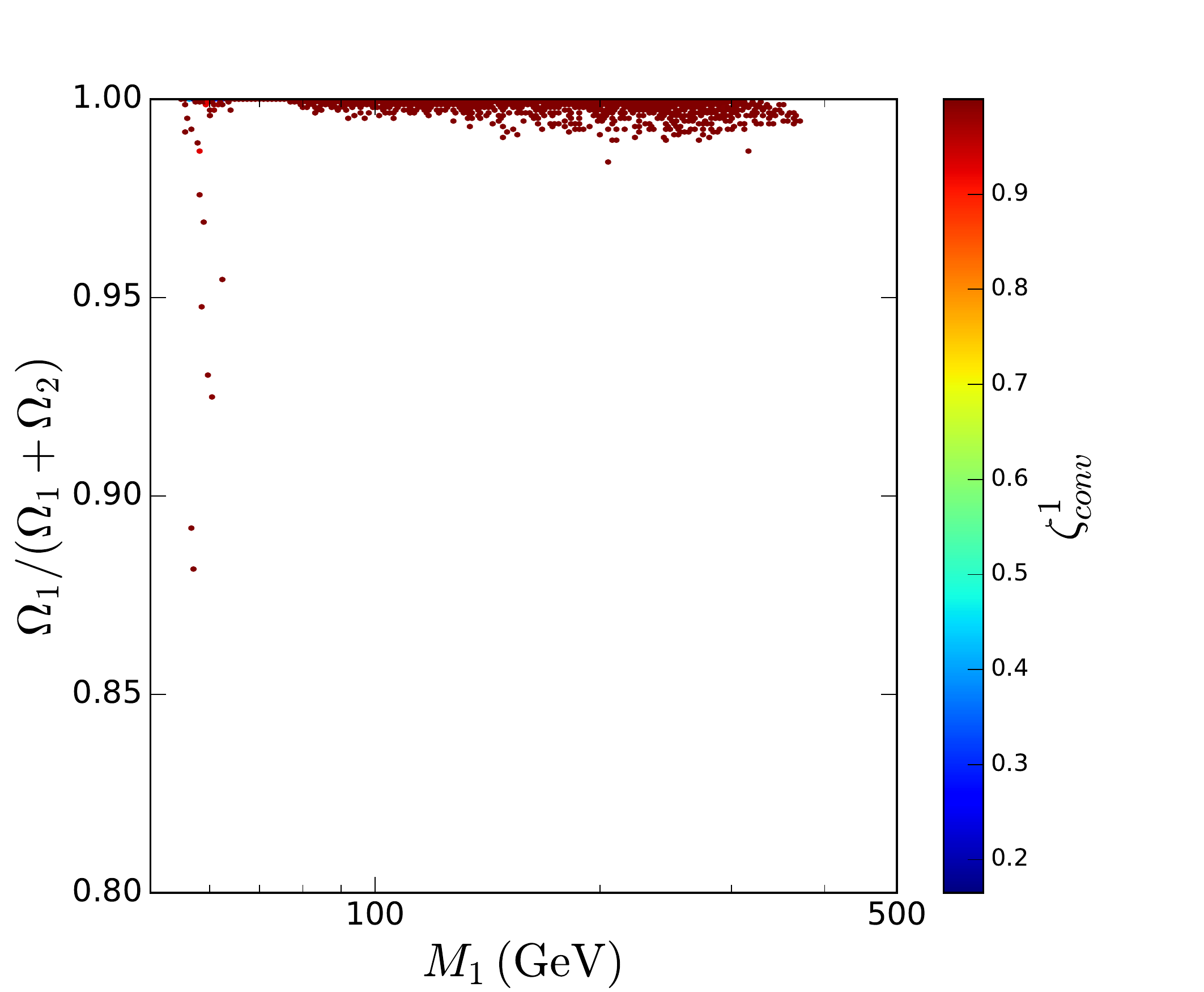}\\
\includegraphics[scale=0.35]{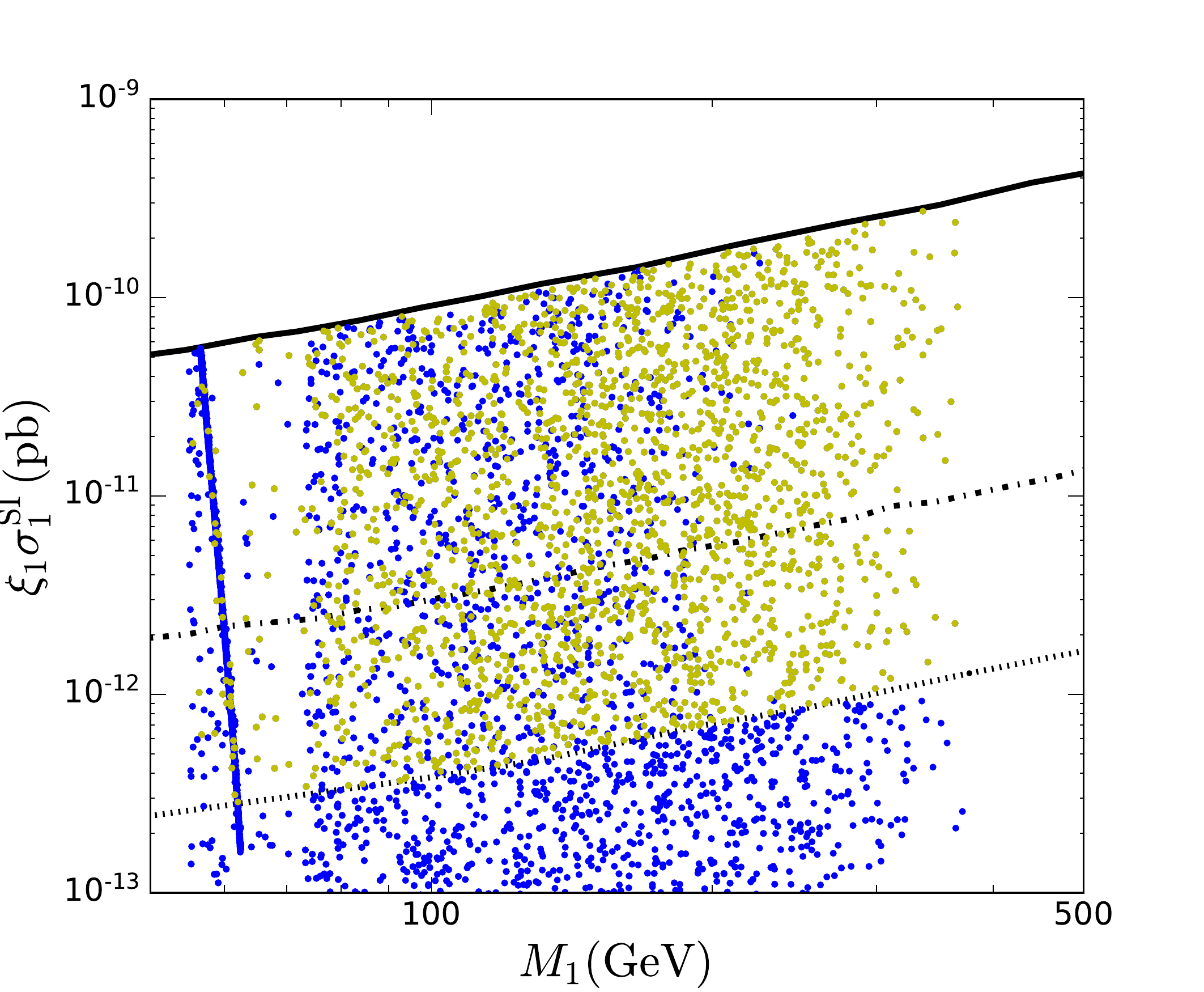}\hspace{0.2cm}
\includegraphics[scale=0.35]{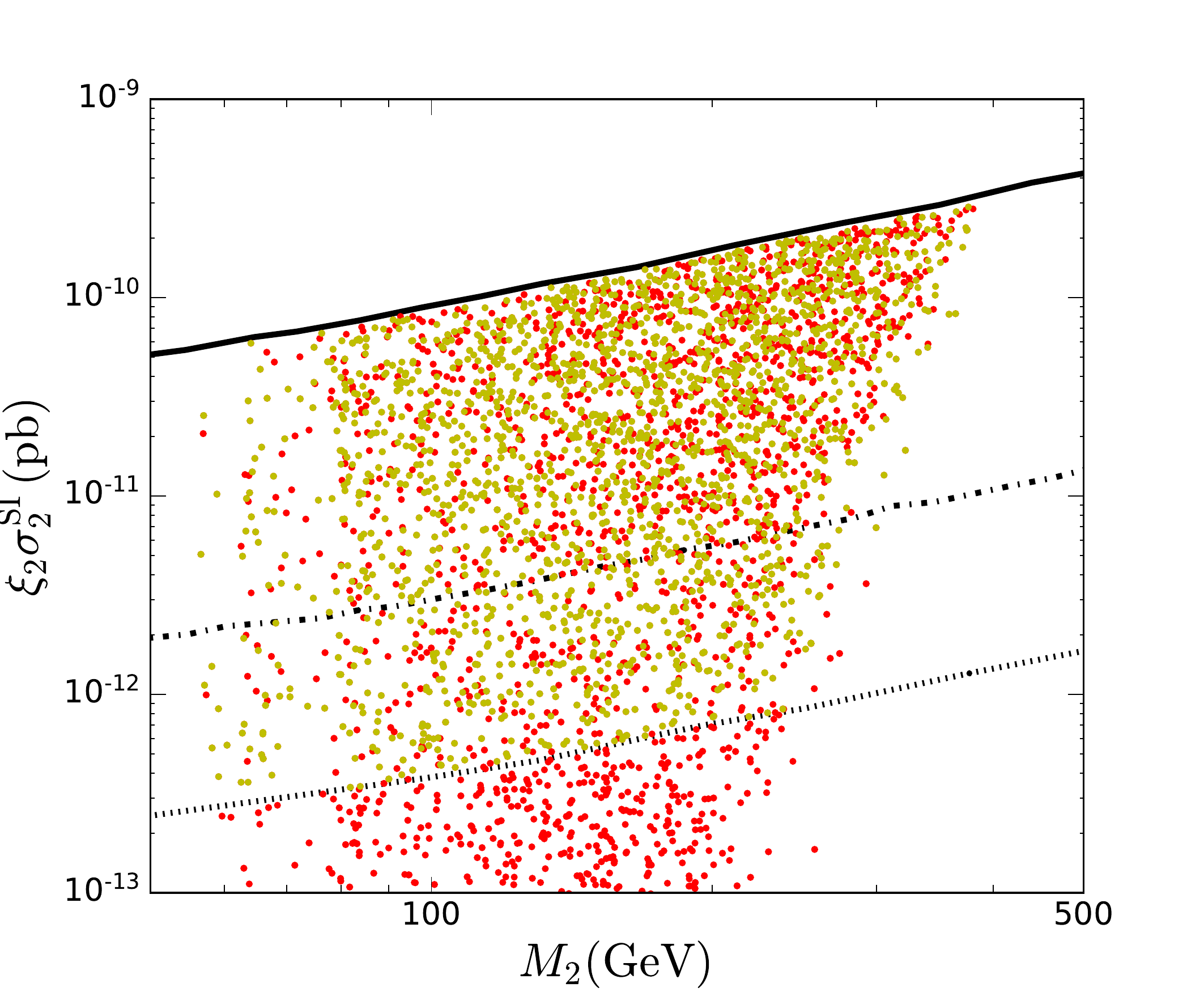}
\caption{\label{fig:scanlamsA} Scan results for $\mu_{Si}=0$ with $\lambda_{3i}\neq0,\, \lambda_{412}\neq0$. Top panels: $M_2/M_1$ (left) and relative contribution of $\phi_1$ to the total DM relic abundance (right) as a function of $M_1$. Bottom panels: spin-independent cross-sections for elastic scattering of $\phi_i$ with nuclei scaled by $\xi_i$. The solid line is the upper limit set by XENON1T collaboration \cite{Aprile:2018dbl} while the dot-dashed and dotted lines show the projected sensitivity of LZ~\cite{Akerib:2018lyp} and DARWIN \cite{Aalbers:2016jon} experiments. Yellow points indicate that both DM particles lay within the sensitivity region of DARWIN.}
\end{figure}

In another scan we allowed  the dimensionless couplings to independently vary  within the range
\begin{align}
    &0.1\leq \lambda_{3i},\lambda_{412}\leq 1. 
\end{align}
while settting $\mu_{Si}=0$. Semi-annihilations are absent in this case so the only new process that can reduce the $\phi_1$ relic density is the conversion $\phi_1+\phi_1\to \phi_1+\phi_2$, which is determined by $\lambda_{31}$ and requires $M_1\sim M_2$ not to be kinematically suppressed during freeze-out. The main results of this scan are displayed in figure \ref{fig:scanlamsA}. From the top-left panel we learn that there is  a new viable region  with  $M_h/2\lesssim M_1\lesssim 400~\mathrm{GeV}$  that is characterized by a high degeneracy between the dark matter particles --$M_2/M_1$ never exceeds $1.1$ there.  As indicated by the value of $\zeta_{conv}^1$, it is the above mentioned conversion process that renders such region consistent with current data. The top-right panel shows that $\phi_1$ essentially accounts for the total dark matter density over the entire new viable region. The contribution of $\phi_2$ amounts to less than $2\%$. In spite of this, either particle could be observed in future direct detection experiments, as illustrated in the bottom panels.

We also did additional scans, including one in which \emph{all} the free parameters of the model are simultaneously varied, and the results are essentially identical to what we found in the three scans already analyzed. It is fair to conclude, therefore, that our scans reveal the genuine viable parameter space of the $Z_5$ model. 

In our analysis so far we have always assumed that $M_1<M_2$ because, as already mentioned in section \ref{sec:model}, the symmetry of the Lagrangian allows us to make this simplification. The results for the case $M_2<M_1$ can be obtained from ours  by simply swapping the corresponding quantities: $M_1\leftrightarrow M_2$, $\mu_{S1}\leftrightarrow \mu_{S2}$, $\lambda_{31}\leftrightarrow \lambda_{32}$, $\Omega_1\leftrightarrow \Omega_2$, etc. Thus, we have actually studied the full range of dark matter masses possible in this model --$M_1/2<M_2<2M_1$.

In this section, the most important results of our work were derived --we characterized the viable parameter space of the $Z_5$ model and determined its detection prospects. Let us review our main findings:
\begin{enumerate}
    \item It is possible to satisfy the relic density constraint and current direct detection limits over the entire range of dark matter masses we considered ($M_1<2$ TeV). In particular, the low mass region $M\lesssim 1$ TeV, which is excluded in the singlet scalar model,  is perfectly compatible with present bounds thanks to the new interactions allowed by the $Z_5$ symmetry.  
    \item The dark matter density is always dominated by the lighter dark matter particle. In our scans, the heavier dark matter particle never accounts for more than $40\%$ of the total density, and often contributes significantly  less than that.
    \item Either dark matter particle may be detected in future direct detection experiments. And in a sizable fraction of models both particles are predicted to be detectable,  providing a way to differentiate this model from the usual scenarios with just one dark matter particle.
\end{enumerate}
Hence, besides being  simple and well-motivated, the $Z_5$ model turns out to be a  consistent and verifiable framework for two-component dark matter.

\section{Discussion}
\label{sec:disc}

We have seen that the new interactions allowed by the $Z_5$ symmetry render this model viable over a wide range of dark matter masses. This result stands in sharp contrast to what is found in similar  models based on $Z_2$ symmetries. In the scenario with  one complex scalar singlet stabilized by a $Z_2$ symmetry,  the dark matter mass necessarily lies either at the  Higgs-resonance  or around 2 TeV, as a consequence of the interplay between the relic density constraint and current direct detection limits. And a similar outcome is obtained in a two-dark matter scenario where the two singlet scalars are stabilized with a $Z_2\times Z_2'$ symmetry.  The $Z_5$ model can be seen as a natural extension of these scenarios and has the advantage of remaining viable at low masses and of being  testable via direct detection experiments.

The $Z_5$ symmetry  used in our model is the lowest $Z_N$ compatible with two dark matter particles that are complex scalar fields \cite{Yaguna:2019cvp}. Even if other $Z_N$  symmetries, with $N>5$, can be imposed to simultaneously stabilize two dark matter particles \cite{Belanger:2014vza}, the $Z_5$ model serves as a prototype for \emph{all} the two-component scenarios  where the dark matter particles are complex scalars. That is, our results can be applied rather straightforwardly to other $Z_N$ frameworks, as explained next.

Let us denote the  two dark matter particles charged under a $Z_N$ by  $\phi_i,\phi_j$ (with $i< j\leq N/2$ and $j\neq N/2$ for $N$ even \cite{Yaguna:2019cvp}), where $\phi_k$ gets a factor $e^{i2\pi k/N}$ upon a $Z_N$ transformation. For $5<N\leq 10$, the  complete set of possibilities for the two dark matter particles is:
\begin{itemize}
\item $(\phi_1,\phi_2)$:  all $Z_N$ symmetries  allow the $\mu_{S1}\phi_1^2\phi_2^*$ term  and forbid the $\mu_{S2}\phi_1\phi_2^2$ and $\lambda_{31}\phi_1^3\phi_2$ terms while the $Z_7$  is the only one that allows $\lambda_{32}\phi_1\phi_2^3$. 
This means that for the scenario with $M_1<M_2$ the viable $M_1$ range can extend up to 2 TeV while for $M_2<M_1$ the maximum value that $M_2$ can reach is 1 TeV.  

\item $(\phi_1,\phi_3)$: the $Z_7$ model allows $\mu_{S3}\phi_3^2\phi_1$ and $\lambda_{31}\phi_1^3\phi_3^*$ which implies a viable mass range up to 2 TeV (1 TeV) for $M_3<M_1$ ($M_1<M_3$). 
For $Z_8$ ($Z_{10}$), only the quartic interactions $\lambda_{31}\phi_1^3\phi_3^*$ and $\lambda_{33}\phi_3^3\phi_1^*$ ($\lambda_{31}\phi_1^3\phi_3^*$ and $\lambda_{33}\phi_3^3\phi_1$) are possible. Consequently, the viable mass range goes up to 400 GeV for both $M_1<M_3$ and $M_3<M_1$ cases. Since $Z_9$ only allows the term $\lambda_{31}\phi_1^3\phi_3^*$ a new viable mass range (up to 400 GeV) is only recovered for $M_1<M_3$. 

\item $(\phi_1,\phi_4)$: $Z_9$ only allows the $\mu_{S4}\phi_4^2\phi_1$ term while $Z_{10}$ forbids all the cubic ($\mu_{Si}$) and quartic $\lambda_{3i}$ interactions. Hence a new viable DM mass range is possible for $Z_9$ models. 

\item $(\phi_2,\phi_3)$: the $Z_7$ model only has $\mu_{S2}\phi_2^2\phi_3$ and $\lambda_{33}\phi_3^3\phi_2^*$ interactions, which imply a viable mass range up to 2 TeV (1 TeV) for $M_2<M_3$ ($M_3<M_2$).
In the $Z_8$ model only the $\mu_{S3}\phi_3^2\phi_2$ term is present such that the viable mass range goes up to 2 TeV (1 TeV) for $M_3<M_2$ ($M_2<M_3$). For $Z_9$ the trilinear interactions are forbidden and only the $\lambda_{32}\phi_2^3\phi_3$ term is allowed. Therefore a new viable mass range (up to 400 GeV) is only recovered for $M_2<M_3$. 
As in the previous item the $Z_{10}$ model forbids both cubic ($\mu_{Si}$) and quartic $\lambda_{3i}$ interactions, which means there is no new viable DM regions. 

\item $(\phi_2,\phi_4)$: the $Z_9$ only allows the $\mu_{S2}\phi_2^2\phi_4^*$ interaction, which implies a viable mass range up to 2 TeV (1 TeV) for $M_2<M_4$ ($M_4<M_2$).
The case of the $Z_{10}$ model is rather special since it features an analogous Lagrangian to the $Z_5$ model which means it allows both cubic ($\mu_{Si}$) and quartic interactions $\lambda_{3i}$. Therefore the results presented in this work apply to the $Z_{10}$ model with $(\phi_2,\phi_4)$ as DM fields. 

\item $(\phi_3,\phi_4)$: the $Z_9$ model only has the $\lambda_{34}\phi_4^3\phi_3^*$ interaction while the $Z_{10}$ model  only allows the $\mu_{S3}\phi_3^2\phi_4$ interaction. It follows that the viable DM mass range goes up to 2 TeV (1 TeV) for $M_3<M_4$ ($M_4<M_3$) in the $Z_{10}$ model, while for $Z_9$ model a new viable mass range (up to 400 GeV) is only recovered for $M_4<M_3$. 
\end{itemize}
This analysis demonstrates that the $Z_5$ model  is the most general $Z_N$ model with two complex fields, from which the DM properties for other models with a higher $Z_N$ symmetry can be deduced to a large extent.  By the same token,   it is  the  $Z_7$ model with $(\phi_1,\phi_2,\phi_3)$ that serves as a prototype for scenarios with three dark matter particles.

Finally, let us comment on possible  extensions of the $Z_5$ model. A simple one is to embed the $Z_5$ symmetry within an spontaneously broken $U(1)$ gauge symmetry \cite{Yaguna:2019cvp}.  In that case, the $\mu_{S1}$ term would still be allowed whereas the $\mu_{S2}$ would require an additional vacuum expectation value. Higher gauge symmetries can also be envisioned.  Another option is to introduce extra fields so as to explain neutrino masses. By including additional vectorlike fermions,  Majorana masses for the neutrinos can be generated at two-loops, as in the $Z_3$-based models studied in ~\cite{Ma:2007gq,Aoki:2014cja,Ding:2016wbd,Ho:2017fte}. The minimal extra fermion content turns out to be two $SU(2)_L$ doublets and one SM singlet, both having the same $Z_5$ charge (either $w_5$ or $w^2_5$) to admit a mixing term via the Higgs doublet.  
It follows that $\phi_1$ and $\phi_2$ become the loop mediators as in the scotogenic models and continue playing the role of DM particles as long as their decays into the new fermions are kinematically closed.  Moreover, in certain regions of parameter space it may be possible to realize a scenario with 3 DM particles (two scalars plus a fermion) without additional symmetries. A phenomenological study of these interesting alternatives lies, however, beyond the scope of the present paper and will be  left for future work.

\section{Conclusions}\label{sec:conc}
We investigated the phenomenology of the two-component dark matter model based on a $Z_5$ symmetry, which serves as an archetype for other $Z_N$ ($N>5$) models with two complex scalar dark matter particles. After describing the model, we studied in detail how  the relic density depends on the new parameters allowed by the $Z_5$ symmetry. In order to characterize the viable parameter space, we did several random scans and analyze their implications. We found that it is possible to satisfy the dark matter constraint and direct detection limits  over the entire range of dark matter masses considered, $M_1\lesssim 2~\mathrm{TeV}$. The key parameter turned out to be the trilinear coupling associated to the lighter dark matter particle (e.g. $\mu_{S1}$ for $M_1<M_2$), which, via semi-annihilations, renders the model viable  without requiring a mass degeneracy between the dark matter particles.  At low dark matter masses ($M_i<1$ TeV), the other trilinear coupling as well as a quartic coupling (e.g. $\mu_{S2}$ and $\lambda_{31}$ for $M_1<M_2$) may also play a role, but only if the dark matter particles are at least mildly degenerate. We found that the dark matter density  is dominated by the lighter particle for all the  viable models and that  a significant fraction of the viable parameter space can be probed by future direct detection experiments. Remarkably, \emph{both} dark matter particles could  give rise to observable signals in such experiments, providing a way not only to test this model but also to differentiate it from  more conventional dark matter scenarios.

\section*{Acknowledgments}
The work of GB and AP was funded by RFBR and CNRS, project number 20-52-15005. COLCIENCIAS supports the work of CY and OZ  through the Grant 111577657253. The work of OZ is further supported by Sostenibilidad-UdeA and the UdeA/CODI Grant 2017-16286. 

\appendix
\section{Scalar potential constraints}
General stability conditions are obtained from copositivity criteria \cite{Kannike:2012pe,Kannike:2016fmd}. For $\lambda_{3i}=0$ they read 
 \begin{align}
 &\lambda_{4i}\geq0,\,\,\,\Lambda_{i}\equiv\lambda_{Si}+2\sqrt{\lambda_{H}\lambda_{4i}}\geq0,\,\,\,\Lambda_3\equiv\lambda_{412}+2\sqrt{\lambda_{41}\lambda_{42}}\geq0,\nonumber\\
 &2\sqrt{\lambda_H\lambda_{41}\lambda_{42}}+\lambda_{S1}\sqrt{\lambda_{42}}+\lambda_{S2}\sqrt{\lambda_{41}}+\lambda_{412}\sqrt{\lambda_{H}}+\sqrt{\Lambda_1\Lambda_2\Lambda_3}\geq0.
 \end{align}
The corresponding expressions for $\lambda_{3i}\neq0$ are rather involved and lengthy. However, taking into account that in our scans the free dimensionless parameters (their absolute values) are at most unity we highlight that the stability conditions may be fulfilled through not so large values for the self-interacting dark matter couplings $\lambda_{4i}$.  
On the other hand, the $Z_5$ symmetry is preserved by requiring $\mu_{i}^2=M_i^2-\lambda_{Si}v_H^2/2>0$. 
\section{RGEs}
The RGEs $dx/d(\ln \mu)=\beta_x^{(1)}/(16\pi^2)$ at one-loop level for the dimensionless scalar parameters are given by
\begin{align} 
\beta_{\lambda_{3i}}^{(1)} & = 6 \lambda_{3i} \Big(2 \lambda_{4i}  + \lambda_{412}\Big),\\ 
\beta_{\lambda_H}^{(1)} & =   \lambda_{S1}^{2} + \lambda_{S2}^{2} + \frac{27}{200} g_{1}^{4} +\frac{9}{20} g_{1}^{2} g_{2}^{2} +\frac{9}{8} g_{2}^{4} -\frac{9}{5} g_{1}^{2} \lambda_H -9 g_{2}^{2} \lambda_H +24 \lambda_H^{2}+12 \lambda_H y_t^2-6y_t^4,\\ 
  \beta_{\lambda_{Si}}^{(1)} & = \left[6y_t^2-\frac{9}{10} g_{1}^{2} -\frac{9}{2} g_{2}^{2}  +12 \lambda_H +8 \lambda_{4i}\right]\lambda_{Si} +4 \lambda_{Si}^{2} +2 \lambda_{412} \lambda_{Sj},\\
\beta_{\lambda_{4i}}^{(1)} & =  
20 \lambda_{4i}^{2}  + 2 \lambda_{Si}^{2}  + \frac{9}{2} |\lambda_{3i}|^2  + \lambda_{412}^{2},\\ 
\beta_{\lambda_{412}}^{(1)} & =  
4 \Big(2 \lambda_{412} \lambda_{42}  + 2 \lambda_{41} \lambda_{412}  + \lambda_{S1} \lambda_{S2}  + \lambda_{412}^{2}\Big) + 9 |\lambda_{31}|^2  + 9 |\lambda_{32}|^2,  
\end{align}
whilst for the dimensionful ones   
\begin{align} 
\beta_{\mu_{S1}}^{(1)} & =  
4( \lambda_{412}  + \lambda_{41}) \mu_{S1}  + 6 \lambda_{31} \mu_{S2}^*  + 6 \lambda_{32} \mu_{S2}, \\ 
\beta_{\mu_{S2}}^{(1)} & =  
4 (\lambda_{412} + \lambda_{42})\mu_{S2}  + 6 \lambda_{31} \mu_{S1}^*  + 6 \mu_{S1} \lambda_{32}^*,\\
\beta_{\mu_H^2}^{(1)} & =  2 \lambda_{S1} \mu_{1}^2  + 2 \lambda_{S2} \mu_{2}^2  -\frac{9}{10} g_{1}^{2} \mu_H^2 -\frac{9}{2} g_{2}^{2} \mu_H^2 +12 \lambda \mu_H^2 +6 \mu_H^2 y_t^2,\\ 
\beta_{\mu_{1}^2}^{(1)} & =  2 \lambda_{412} \mu_{2}^2  + 2 |\mu_{S1}|^2  + 4 \lambda_{S1} \mu_H^2  + 8 \lambda_{41} \mu_{1}^2  + |\mu_{S2}|^2,\\ 
\beta_{\mu_{2}^2}^{(1)} & =  2 \lambda_{412} \mu_{1}^2  + 2 |\mu_{S2}|^2  + 4 \lambda_{S2} \mu_H^2  + 8 \lambda_{42} \mu_{2}^2  + |\mu_{S1}|^2. 
\end{align}
These analytical expressions were derived by implementing the model in {\tt SARAH-4.12.3}~\cite{Staub:2008uz,Staub:2013tta}.

\bibliographystyle{apsrev4-1long}
\bibliography{references}

\begin{thebibliography}{10}%
\makeatletter
\providecommand \@ifxundefined [1]{%
 \ifx #1\undefined \expandafter \@firstoftwo
 \else \expandafter \@secondoftwo
\fi
}%
\providecommand \@ifnum [1]{%
 \ifnum #1\expandafter \@firstoftwo
 \else \expandafter \@secondoftwo
\fi
}%
\providecommand \enquote [1]{``#1''}%
\providecommand \bibnamefont  [1]{#1}%
\providecommand \bibfnamefont [1]{#1}%
\providecommand \citenamefont [1]{#1}%
\providecommand\href[0]{\@sanitize\@href}%
\providecommand\@href[1]{\endgroup\@@startlink{#1}\endgroup\@@href}%
\providecommand\@@href[1]{#1\@@endlink}%
\providecommand \@sanitize [0]{\begingroup\catcode`\&12\catcode`\#12\relax}%
\@ifxundefined \pdfoutput {\@firstoftwo}{%
 \@ifnum{\z@=\pdfoutput}{\@firstoftwo}{\@secondoftwo}%
}{%
 \providecommand\@@startlink[1]{\leavevmode\special{html:<a href="#1">}}%
 \providecommand\@@endlink[0]{\special{html:</a>}}%
}{%
 \providecommand\@@startlink[1]{%
  \leavevmode
  \pdfstartlink
   attr{/Border[0 0 1 ]/H/I/C[0 1 1]}%
   user{/Subtype/Link/A<</Type/Action/S/URI/URI(#1)>>}%
  \relax
 }%
 \providecommand\@@endlink[0]{\pdfendlink}%
}%
\providecommand \url  [0]{\begingroup\@sanitize \@url }%
\providecommand \@url [1]{\endgroup\@href {#1}{\urlprefix}}%
\providecommand \urlprefix [0]{URL }%
\providecommand \Eprint[0]{\href }%
\@ifxundefined \urlstyle {%
  \providecommand \doi [1]{doi:\discretionary{}{}{}#1}%
}{%
  \providecommand \doi [0]{doi:\discretionary{}{}{}\begingroup
  \urlstyle{rm}\Url }%
}%
\providecommand \doibase [0]{http://dx.doi.org/}%
\providecommand \Doi[1]{\href{\doibase#1}}%
\providecommand \bibAnnote [3]{%
  \BibitemShut{#1}%
  \begin{quotation}\noindent
    \textsc{Key:}\ #2\\\textsc{Annotation:}\ #3%
  \end{quotation}%
}%
\providecommand \bibAnnoteFile [2]{%
  \IfFileExists{#2}{\bibAnnote {#1} {#2} {\input{#2}}}{}%
}%
\providecommand \typeout [0]{\immediate \write \m@ne }%
\providecommand \selectlanguage [0]{\@gobble}%
\providecommand \bibinfo [0]{\@secondoftwo}%
\providecommand \bibfield [0]{\@secondoftwo}%
\providecommand \translation [1]{[#1]}%
\providecommand \BibitemOpen[0]{}%
\providecommand \bibitemStop [0]{}%
\providecommand \bibitemNoStop [0]{.\EOS\space}%
\providecommand \EOS [0]{\spacefactor3000\relax}%
\providecommand \BibitemShut [1]{\csname bibitem#1\endcsname}%
\bibitem{Boehm:2003ha}%
  \BibitemOpen
  \bibfield{author}{%
  \bibinfo {author} {\bibfnamefont{C.}~\bibnamefont{Boehm}}, \bibinfo {author}
  {\bibfnamefont{Pierre}\ \bibnamefont{Fayet}},\ and\ \bibinfo {author}
  {\bibfnamefont{J.}~\bibnamefont{Silk}},\ }%
  \bibfield{title}{%
  \enquote{\bibinfo {title} {{Light and heavy dark matter particles}},}\ }%
  \bibfield{journal}{%
  \Doi{10.1103/PhysRevD.69.101302}{\bibinfo {journal} {Phys. Rev.}}\ }%
  \textbf{\bibinfo {volume} {D69}},\ \bibinfo {pages} {101302} (\bibinfo {year}
  {2004}),\ \Eprint{http://arxiv.org/abs/hep-ph/0311143}{arXiv:hep-ph/0311143
  [hep-ph]}%
  \bibAnnoteFile{NoStop}{Boehm:2003ha}%
\bibitem{Ma:2006uv}%
  \BibitemOpen
  \bibfield{author}{%
  \bibinfo {author} {\bibfnamefont{Ernest}\ \bibnamefont{Ma}},\ }%
  \bibfield{title}{%
  \enquote{\bibinfo {title} {{Supersymmetric Model of Radiative Seesaw Majorana
  Neutrino Masses}},}\ }%
  \bibfield{journal}{%
  \bibinfo {journal} {Annales Fond. Broglie}\ }%
  \textbf{\bibinfo {volume} {31}},\ \bibinfo {pages} {285} (\bibinfo {year}
  {2006}),\ \Eprint{http://arxiv.org/abs/hep-ph/0607142}{arXiv:hep-ph/0607142
  [hep-ph]}%
  \bibAnnoteFile{NoStop}{Ma:2006uv}%
\bibitem{Cao:2007fy}%
  \BibitemOpen
  \bibfield{author}{%
  \bibinfo {author} {\bibfnamefont{Qing-Hong}\ \bibnamefont{Cao}}, \bibinfo
  {author} {\bibfnamefont{Ernest}\ \bibnamefont{Ma}}, \bibinfo {author}
  {\bibfnamefont{Jose}\ \bibnamefont{Wudka}},\ and\ \bibinfo {author}
  {\bibfnamefont{C.~P.}\ \bibnamefont{Yuan}},\ }%
  \bibfield{title}{%
  \enquote{\bibinfo {title} {{Multipartite dark matter}},}\ }%
   (\bibinfo {year} {2007}),\
  \Eprint{http://arxiv.org/abs/0711.3881}{arXiv:0711.3881 [hep-ph]}%
  \bibAnnoteFile{NoStop}{Cao:2007fy}%
\bibitem{Hur:2007ur}%
  \BibitemOpen
  \bibfield{author}{%
  \bibinfo {author} {\bibfnamefont{Taeil}\ \bibnamefont{Hur}}, \bibinfo
  {author} {\bibfnamefont{Hye-Sung}\ \bibnamefont{Lee}},\ and\ \bibinfo
  {author} {\bibfnamefont{Salah}\ \bibnamefont{Nasri}},\ }%
  \bibfield{title}{%
  \enquote{\bibinfo {title} {{A Supersymmetric U(1)-prime model with multiple
  dark matters}},}\ }%
  \bibfield{journal}{%
  \Doi{10.1103/PhysRevD.77.015008}{\bibinfo {journal} {Phys. Rev.}}\ }%
  \textbf{\bibinfo {volume} {D77}},\ \bibinfo {pages} {015008} (\bibinfo {year}
  {2008}),\ \Eprint{http://arxiv.org/abs/0710.2653}{arXiv:0710.2653 [hep-ph]}%
  \bibAnnoteFile{NoStop}{Hur:2007ur}%
\bibitem{Lee:2008pc}%
  \BibitemOpen
  \bibfield{author}{%
  \bibinfo {author} {\bibfnamefont{Hye-Sung}\ \bibnamefont{Lee}},\ }%
  \bibfield{title}{%
  \enquote{\bibinfo {title} {{Lightest U-parity Particle (LUP) dark matter}},}\
  }%
  \bibfield{journal}{%
  \Doi{10.1016/j.physletb.2008.03.065}{\bibinfo {journal} {Phys. Lett.}}\ }%
  \textbf{\bibinfo {volume} {B663}},\ \bibinfo {pages} {255--258} (\bibinfo
  {year} {2008}),\ \Eprint{http://arxiv.org/abs/0802.0506}{arXiv:0802.0506
  [hep-ph]}%
  \bibAnnoteFile{NoStop}{Lee:2008pc}%
\bibitem{Zurek:2008qg}%
  \BibitemOpen
  \bibfield{author}{%
  \bibinfo {author} {\bibfnamefont{Kathryn~M.}\ \bibnamefont{Zurek}},\ }%
  \bibfield{title}{%
  \enquote{\bibinfo {title} {{Multi-Component Dark Matter}},}\ }%
  \bibfield{journal}{%
  \Doi{10.1103/PhysRevD.79.115002}{\bibinfo {journal} {Phys. Rev.}}\ }%
  \textbf{\bibinfo {volume} {D79}},\ \bibinfo {pages} {115002} (\bibinfo {year}
  {2009}),\ \Eprint{http://arxiv.org/abs/0811.4429}{arXiv:0811.4429 [hep-ph]}%
  \bibAnnoteFile{NoStop}{Zurek:2008qg}%
\bibitem{Profumo:2009tb}%
  \BibitemOpen
  \bibfield{author}{%
  \bibinfo {author} {\bibfnamefont{Stefano}\ \bibnamefont{Profumo}}, \bibinfo
  {author} {\bibfnamefont{Kris}\ \bibnamefont{Sigurdson}},\ and\ \bibinfo
  {author} {\bibfnamefont{Lorenzo}\ \bibnamefont{Ubaldi}},\ }%
  \bibfield{title}{%
  \enquote{\bibinfo {title} {{Can we discover multi-component WIMP dark
  matter?}}.}\ }%
  \bibfield{journal}{%
  \Doi{10.1088/1475-7516/2009/12/016}{\bibinfo {journal} {JCAP}}\ }%
  \textbf{\bibinfo {volume} {0912}},\ \bibinfo {pages} {016} (\bibinfo {year}
  {2009}),\ \Eprint{http://arxiv.org/abs/0907.4374}{arXiv:0907.4374 [hep-ph]}%
  \bibAnnoteFile{NoStop}{Profumo:2009tb}%
\bibitem{Baer:2011hx}%
  \BibitemOpen
  \bibfield{author}{%
  \bibinfo {author} {\bibfnamefont{Howard}\ \bibnamefont{Baer}}, \bibinfo
  {author} {\bibfnamefont{Andre}\ \bibnamefont{Lessa}}, \bibinfo {author}
  {\bibfnamefont{Shibi}\ \bibnamefont{Rajagopalan}},\ and\ \bibinfo {author}
  {\bibfnamefont{Warintorn}\ \bibnamefont{Sreethawong}},\ }%
  \bibfield{title}{%
  \enquote{\bibinfo {title} {{Mixed axion/neutralino cold dark matter in
  supersymmetric models}},}\ }%
  \bibfield{journal}{%
  \Doi{10.1088/1475-7516/2011/06/031}{\bibinfo {journal} {JCAP}}\ }%
  \textbf{\bibinfo {volume} {1106}},\ \bibinfo {pages} {031} (\bibinfo {year}
  {2011}),\ \Eprint{http://arxiv.org/abs/1103.5413}{arXiv:1103.5413 [hep-ph]}%
  \bibAnnoteFile{NoStop}{Baer:2011hx}%
\bibitem{Esch:2014jpa}%
  \BibitemOpen
  \bibfield{author}{%
  \bibinfo {author} {\bibfnamefont{Sonja}\ \bibnamefont{Esch}}, \bibinfo
  {author} {\bibfnamefont{Michael}\ \bibnamefont{Klasen}},\ and\ \bibinfo
  {author} {\bibfnamefont{Carlos~E.}\ \bibnamefont{Yaguna}},\ }%
  \bibfield{title}{%
  \enquote{\bibinfo {title} {{A minimal model for two-component dark
  matter}},}\ }%
  \bibfield{journal}{%
  \Doi{10.1007/JHEP09(2014)108}{\bibinfo {journal} {JHEP}}\ }%
  \textbf{\bibinfo {volume} {09}},\ \bibinfo {pages} {108} (\bibinfo {year}
  {2014}),\ \Eprint{http://arxiv.org/abs/1406.0617}{arXiv:1406.0617 [hep-ph]}%
  \bibAnnoteFile{NoStop}{Esch:2014jpa}%
\bibitem{Batell:2010bp}%
  \BibitemOpen
  \bibfield{author}{%
  \bibinfo {author} {\bibfnamefont{Brian}\ \bibnamefont{Batell}},\ }%
  \bibfield{title}{%
  \enquote{\bibinfo {title} {{Dark Discrete Gauge Symmetries}},}\ }%
  \bibfield{journal}{%
  \Doi{10.1103/PhysRevD.83.035006}{\bibinfo {journal} {Phys. Rev.}}\ }%
  \textbf{\bibinfo {volume} {D83}},\ \bibinfo {pages} {035006} (\bibinfo {year}
  {2011}),\ \Eprint{http://arxiv.org/abs/1007.0045}{arXiv:1007.0045 [hep-ph]}%
  \bibAnnoteFile{NoStop}{Batell:2010bp}%
\bibitem{Belanger:2014bga}%
  \BibitemOpen
  \bibfield{author}{%
  \bibinfo {author} {\bibfnamefont{Geneviève}\ \bibnamefont{Bélanger}},
  \bibinfo {author} {\bibfnamefont{Kristjan}\ \bibnamefont{Kannike}}, \bibinfo
  {author} {\bibfnamefont{Alexander}\ \bibnamefont{Pukhov}},\ and\ \bibinfo
  {author} {\bibfnamefont{Martti}\ \bibnamefont{Raidal}},\ }%
  \bibfield{title}{%
  \enquote{\bibinfo {title} {{Minimal semi-annihilating $\mathbb{Z}_N$ scalar
  dark matter}},}\ }%
  \bibfield{journal}{%
  \Doi{10.1088/1475-7516/2014/06/021}{\bibinfo {journal} {JCAP}}\ }%
  \textbf{\bibinfo {volume} {1406}},\ \bibinfo {pages} {021} (\bibinfo {year}
  {2014}),\ \Eprint{http://arxiv.org/abs/1403.4960}{arXiv:1403.4960 [hep-ph]}%
  \bibAnnoteFile{NoStop}{Belanger:2014bga}%
\bibitem{Yaguna:2019cvp}%
  \BibitemOpen
  \bibfield{author}{%
  \bibinfo {author} {\bibfnamefont{Carlos~E.}\ \bibnamefont{Yaguna}}\ and\
  \bibinfo {author} {\bibfnamefont{Óscar}\ \bibnamefont{Zapata}},\ }%
  \bibfield{title}{%
  \enquote{\bibinfo {title} {{Multi-component scalar dark matter from a $Z_N$
  symmetry: a systematic analysis}},}\ }%
  \bibfield{journal}{%
  \Doi{10.1007/JHEP03(2020)109}{\bibinfo {journal} {JHEP}}\ }%
  \textbf{\bibinfo {volume} {03}},\ \bibinfo {pages} {109} (\bibinfo {year}
  {2020}),\ \Eprint{http://arxiv.org/abs/1911.05515}{arXiv:1911.05515
  [hep-ph]}%
  \bibAnnoteFile{NoStop}{Yaguna:2019cvp}%
\bibitem{Belanger:2014vza}%
  \BibitemOpen
  \bibfield{author}{%
  \bibinfo {author} {\bibfnamefont{G.}~\bibnamefont{Bélanger}}, \bibinfo
  {author} {\bibfnamefont{F.}~\bibnamefont{Boudjema}}, \bibinfo {author}
  {\bibfnamefont{A.}~\bibnamefont{Pukhov}},\ and\ \bibinfo {author}
  {\bibfnamefont{A.}~\bibnamefont{Semenov}},\ }%
  \bibfield{title}{%
  \enquote{\bibinfo {title} {{micrOMEGAs4.1: two dark matter candidates}},}\ }%
  \bibfield{journal}{%
  \Doi{10.1016/j.cpc.2015.03.003}{\bibinfo {journal} {Comput. Phys. Commun.}}\
  }%
  \textbf{\bibinfo {volume} {192}},\ \bibinfo {pages} {322--329} (\bibinfo
  {year} {2015}),\ \Eprint{http://arxiv.org/abs/1407.6129}{arXiv:1407.6129
  [hep-ph]}%
  \bibAnnoteFile{NoStop}{Belanger:2014vza}%
\bibitem{Arcadi:2019lka}%
  \BibitemOpen
  \bibfield{author}{%
  \bibinfo {author} {\bibfnamefont{Giorgio}\ \bibnamefont{Arcadi}}, \bibinfo
  {author} {\bibfnamefont{Abdelhak}\ \bibnamefont{Djouadi}},\ and\ \bibinfo
  {author} {\bibfnamefont{Martti}\ \bibnamefont{Raidal}},\ }%
  \bibfield{title}{%
  \enquote{\bibinfo {title} {{Dark Matter through the Higgs portal}},}\ }%
  \bibfield{journal}{%
  \Doi{10.1016/j.physrep.2019.11.003}{\bibinfo {journal} {Phys. Rept.}}\ }%
  \textbf{\bibinfo {volume} {842}},\ \bibinfo {pages} {1--180} (\bibinfo {year}
  {2020}),\ \Eprint{http://arxiv.org/abs/1903.03616}{arXiv:1903.03616
  [hep-ph]}%
  \bibAnnoteFile{NoStop}{Arcadi:2019lka}%
\bibitem{Silveira:1985rk}%
  \BibitemOpen
  \bibfield{author}{%
  \bibinfo {author} {\bibfnamefont{Vanda}\ \bibnamefont{Silveira}}\ and\
  \bibinfo {author} {\bibfnamefont{A.}~\bibnamefont{Zee}},\ }%
  \bibfield{title}{%
  \enquote{\bibinfo {title} {{SCALAR PHANTOMS}},}\ }%
  \bibfield{journal}{%
  \Doi{10.1016/0370-2693(85)90624-0}{\bibinfo {journal} {Phys. Lett. B}}\ }%
  \textbf{\bibinfo {volume} {161}},\ \bibinfo {pages} {136--140} (\bibinfo
  {year} {1985})%
  \bibAnnoteFile{NoStop}{Silveira:1985rk}%
\bibitem{McDonald:1993ex}%
  \BibitemOpen
  \bibfield{author}{%
  \bibinfo {author} {\bibfnamefont{John}\ \bibnamefont{McDonald}},\ }%
  \bibfield{title}{%
  \enquote{\bibinfo {title} {{Gauge singlet scalars as cold dark matter}},}\ }%
  \bibfield{journal}{%
  \Doi{10.1103/PhysRevD.50.3637}{\bibinfo {journal} {Phys. Rev.}}\ }%
  \textbf{\bibinfo {volume} {D50}},\ \bibinfo {pages} {3637--3649} (\bibinfo
  {year} {1994}),\
  \Eprint{http://arxiv.org/abs/hep-ph/0702143}{arXiv:hep-ph/0702143 [HEP-PH]}%
  \bibAnnoteFile{NoStop}{McDonald:1993ex}%
\bibitem{Burgess:2000yq}%
  \BibitemOpen
  \bibfield{author}{%
  \bibinfo {author} {\bibfnamefont{C.P.}\ \bibnamefont{Burgess}}, \bibinfo
  {author} {\bibfnamefont{Maxim}\ \bibnamefont{Pospelov}},\ and\ \bibinfo
  {author} {\bibfnamefont{Tonnis}\ \bibnamefont{ter Veldhuis}},\ }%
  \bibfield{title}{%
  \enquote{\bibinfo {title} {{The Minimal model of nonbaryonic dark matter: A
  Singlet scalar}},}\ }%
  \bibfield{journal}{%
  \Doi{10.1016/S0550-3213(01)00513-2}{\bibinfo {journal} {Nucl. Phys. B}}\ }%
  \textbf{\bibinfo {volume} {619}},\ \bibinfo {pages} {709--728} (\bibinfo
  {year} {2001}),\
  \Eprint{http://arxiv.org/abs/hep-ph/0011335}{arXiv:hep-ph/0011335}%
  \bibAnnoteFile{NoStop}{Burgess:2000yq}%
\bibitem{Cline:2013gha}%
  \BibitemOpen
  \bibfield{author}{%
  \bibinfo {author} {\bibfnamefont{James~M.}\ \bibnamefont{Cline}}, \bibinfo
  {author} {\bibfnamefont{Kimmo}\ \bibnamefont{Kainulainen}}, \bibinfo {author}
  {\bibfnamefont{Pat}\ \bibnamefont{Scott}},\ and\ \bibinfo {author}
  {\bibfnamefont{Christoph}\ \bibnamefont{Weniger}},\ }%
  \bibfield{title}{%
  \enquote{\bibinfo {title} {{Update on scalar singlet dark matter}},}\ }%
  \bibfield{journal}{%
  \Doi{10.1103/PhysRevD.92.039906, 10.1103/PhysRevD.88.055025}{\bibinfo
  {journal} {Phys. Rev.}}\ }%
  \textbf{\bibinfo {volume} {D88}},\ \bibinfo {pages} {055025} (\bibinfo {year}
  {2013}),\ \bibinfo {note} {[Erratum: Phys. Rev.D92,no.3,039906(2015)]},\
  \Eprint{http://arxiv.org/abs/1306.4710}{arXiv:1306.4710 [hep-ph]}%
  \bibAnnoteFile{NoStop}{Cline:2013gha}%
\bibitem{Athron:2018ipf}%
  \BibitemOpen
  \bibfield{author}{%
  \bibinfo {author} {\bibfnamefont{Peter}\ \bibnamefont{Athron}}, \bibinfo
  {author} {\bibfnamefont{Jonathan~M.}\ \bibnamefont{Cornell}}, \bibinfo
  {author} {\bibfnamefont{Felix}\ \bibnamefont{Kahlhoefer}}, \bibinfo {author}
  {\bibfnamefont{James}\ \bibnamefont{Mckay}}, \bibinfo {author}
  {\bibfnamefont{Pat}\ \bibnamefont{Scott}},\ and\ \bibinfo {author}
  {\bibfnamefont{Sebastian}\ \bibnamefont{Wild}},\ }%
  \bibfield{title}{%
  \enquote{\bibinfo {title} {{Impact of vacuum stability, perturbativity and
  XENON1T on global fits of $\mathbb {Z}_2$ and $\mathbb {Z}_3$ scalar singlet
  dark matter}},}\ }%
  \bibfield{journal}{%
  \Doi{10.1140/epjc/s10052-018-6314-y}{\bibinfo {journal} {Eur. Phys. J. C}}\
  }%
  \textbf{\bibinfo {volume} {78}},\ \bibinfo {pages} {830} (\bibinfo {year}
  {2018}),\ \Eprint{http://arxiv.org/abs/1806.11281}{arXiv:1806.11281
  [hep-ph]}%
  \bibAnnoteFile{NoStop}{Athron:2018ipf}%
\bibitem{Hall:2009bx}%
  \BibitemOpen
  \bibfield{author}{%
  \bibinfo {author} {\bibfnamefont{Lawrence~J.}\ \bibnamefont{Hall}}, \bibinfo
  {author} {\bibfnamefont{Karsten}\ \bibnamefont{Jedamzik}}, \bibinfo {author}
  {\bibfnamefont{John}\ \bibnamefont{March-Russell}},\ and\ \bibinfo {author}
  {\bibfnamefont{Stephen~M.}\ \bibnamefont{West}},\ }%
  \bibfield{title}{%
  \enquote{\bibinfo {title} {{Freeze-In Production of FIMP Dark Matter}},}\ }%
  \bibfield{journal}{%
  \Doi{10.1007/JHEP03(2010)080}{\bibinfo {journal} {JHEP}}\ }%
  \textbf{\bibinfo {volume} {03}},\ \bibinfo {pages} {080} (\bibinfo {year}
  {2010}),\ \Eprint{http://arxiv.org/abs/0911.1120}{arXiv:0911.1120 [hep-ph]}%
  \bibAnnoteFile{NoStop}{Hall:2009bx}%
\bibitem{Yaguna:2011qn}%
  \BibitemOpen
  \bibfield{author}{%
  \bibinfo {author} {\bibfnamefont{Carlos~E.}\ \bibnamefont{Yaguna}},\ }%
  \bibfield{title}{%
  \enquote{\bibinfo {title} {{The Singlet Scalar as FIMP Dark Matter}},}\ }%
  \bibfield{journal}{%
  \Doi{10.1007/JHEP08(2011)060}{\bibinfo {journal} {JHEP}}\ }%
  \textbf{\bibinfo {volume} {08}},\ \bibinfo {pages} {060} (\bibinfo {year}
  {2011}),\ \Eprint{http://arxiv.org/abs/1105.1654}{arXiv:1105.1654 [hep-ph]}%
  \bibAnnoteFile{NoStop}{Yaguna:2011qn}%
\bibitem{DEramo:2010keq}%
  \BibitemOpen
  \bibfield{author}{%
  \bibinfo {author} {\bibfnamefont{Francesco}\ \bibnamefont{D'Eramo}}\ and\
  \bibinfo {author} {\bibfnamefont{Jesse}\ \bibnamefont{Thaler}},\ }%
  \bibfield{title}{%
  \enquote{\bibinfo {title} {{Semi-annihilation of Dark Matter}},}\ }%
  \bibfield{journal}{%
  \Doi{10.1007/JHEP06(2010)109}{\bibinfo {journal} {JHEP}}\ }%
  \textbf{\bibinfo {volume} {06}},\ \bibinfo {pages} {109} (\bibinfo {year}
  {2010}),\ \Eprint{http://arxiv.org/abs/1003.5912}{arXiv:1003.5912 [hep-ph]}%
  \bibAnnoteFile{NoStop}{DEramo:2010keq}%
\bibitem{Belanger:2013oya}%
  \BibitemOpen
  \bibfield{author}{%
  \bibinfo {author} {\bibfnamefont{G.}~\bibnamefont{Belanger}}, \bibinfo
  {author} {\bibfnamefont{F.}~\bibnamefont{Boudjema}}, \bibinfo {author}
  {\bibfnamefont{A.}~\bibnamefont{Pukhov}},\ and\ \bibinfo {author}
  {\bibfnamefont{A.}~\bibnamefont{Semenov}},\ }%
  \bibfield{title}{%
  \enquote{\bibinfo {title} {{micrOMEGAs\_3: A program for calculating dark
  matter observables}},}\ }%
  \bibfield{journal}{%
  \Doi{10.1016/j.cpc.2013.10.016}{\bibinfo {journal} {Comput. Phys. Commun.}}\
  }%
  \textbf{\bibinfo {volume} {185}},\ \bibinfo {pages} {960--985} (\bibinfo
  {year} {2014}),\ \Eprint{http://arxiv.org/abs/1305.0237}{arXiv:1305.0237
  [hep-ph]}%
  \bibAnnoteFile{NoStop}{Belanger:2013oya}%
\bibitem{Belanger:2018ccd}%
  \BibitemOpen
  \bibfield{author}{%
  \bibinfo {author} {\bibfnamefont{Geneviève}\ \bibnamefont{Bélanger}},
  \bibinfo {author} {\bibfnamefont{Fawzi}\ \bibnamefont{Boudjema}}, \bibinfo
  {author} {\bibfnamefont{Andreas}\ \bibnamefont{Goudelis}}, \bibinfo {author}
  {\bibfnamefont{Alexander}\ \bibnamefont{Pukhov}},\ and\ \bibinfo {author}
  {\bibfnamefont{Bryan}\ \bibnamefont{Zaldivar}},\ }%
  \bibfield{title}{%
  \enquote{\bibinfo {title} {{micrOMEGAs5.0 : Freeze-in}},}\ }%
  \bibfield{journal}{%
  \Doi{10.1016/j.cpc.2018.04.027}{\bibinfo {journal} {Comput. Phys. Commun.}}\
  }%
  \textbf{\bibinfo {volume} {231}},\ \bibinfo {pages} {173--186} (\bibinfo
  {year} {2018}),\ \Eprint{http://arxiv.org/abs/1801.03509}{arXiv:1801.03509
  [hep-ph]}%
  \bibAnnoteFile{NoStop}{Belanger:2018ccd}%
\bibitem{Aprile:2018dbl}%
  \BibitemOpen
  \bibfield{author}{%
  \bibinfo {author} {\bibfnamefont{E.}~\bibnamefont{Aprile}} \emph{et~al.}
  (\bibinfo {collaboration} {XENON}),\ }%
  \bibfield{title}{%
  \enquote{\bibinfo {title} {{Dark Matter Search Results from a One
  Tonne$\times$Year Exposure of XENON1T}},}\ }%
   (\bibinfo {year} {2018}),\
  \Eprint{http://arxiv.org/abs/1805.12562}{arXiv:1805.12562 [astro-ph.CO]}%
  \bibAnnoteFile{NoStop}{Aprile:2018dbl}%
\bibitem{Akerib:2018lyp}%
  \BibitemOpen
  \bibfield{author}{%
  \bibinfo {author} {\bibfnamefont{D.~S.}\ \bibnamefont{Akerib}} \emph{et~al.}
  (\bibinfo {collaboration} {LUX-ZEPLIN}),\ }%
  \bibfield{title}{%
  \enquote{\bibinfo {title} {{Projected WIMP sensitivity of the LUX-ZEPLIN (LZ)
  dark matter experiment}},}\ }%
   (\bibinfo {year} {2018}),\
  \Eprint{http://arxiv.org/abs/1802.06039}{arXiv:1802.06039 [astro-ph.IM]}%
  \bibAnnoteFile{NoStop}{Akerib:2018lyp}%
\bibitem{Aalbers:2016jon}%
  \BibitemOpen
  \bibfield{author}{%
  \bibinfo {author} {\bibfnamefont{J.}~\bibnamefont{Aalbers}} \emph{et~al.}
  (\bibinfo {collaboration} {DARWIN}),\ }%
  \bibfield{title}{%
  \enquote{\bibinfo {title} {{DARWIN: towards the ultimate dark matter
  detector}},}\ }%
  \bibfield{journal}{%
  \Doi{10.1088/1475-7516/2016/11/017}{\bibinfo {journal} {JCAP}}\ }%
  \textbf{\bibinfo {volume} {1611}},\ \bibinfo {pages} {017} (\bibinfo {year}
  {2016}),\ \Eprint{http://arxiv.org/abs/1606.07001}{arXiv:1606.07001
  [astro-ph.IM]}%
  \bibAnnoteFile{NoStop}{Aalbers:2016jon}%
\bibitem{Ackermann:2015zua}%
  \BibitemOpen
  \bibfield{author}{%
  \bibinfo {author} {\bibfnamefont{M.}~\bibnamefont{Ackermann}} \emph{et~al.}
  (\bibinfo {collaboration} {Fermi-LAT}),\ }%
  \bibfield{title}{%
  \enquote{\bibinfo {title} {{Searching for Dark Matter Annihilation from Milky
  Way Dwarf Spheroidal Galaxies with Six Years of Fermi Large Area Telescope
  Data}},}\ }%
  \bibfield{journal}{%
  \Doi{10.1103/PhysRevLett.115.231301}{\bibinfo {journal} {Phys. Rev. Lett.}}\
  }%
  \textbf{\bibinfo {volume} {115}},\ \bibinfo {pages} {231301} (\bibinfo {year}
  {2015}),\ \Eprint{http://arxiv.org/abs/1503.02641}{arXiv:1503.02641
  [astro-ph.HE]}%
  \bibAnnoteFile{NoStop}{Ackermann:2015zua}%
\bibitem{Charles:2016pgz}%
  \BibitemOpen
  \bibfield{author}{%
  \bibinfo {author} {\bibfnamefont{E.}~\bibnamefont{Charles}} \emph{et~al.}
  (\bibinfo {collaboration} {Fermi-LAT}),\ }%
  \bibfield{title}{%
  \enquote{\bibinfo {title} {{Sensitivity Projections for Dark Matter Searches
  with the Fermi Large Area Telescope}},}\ }%
  \bibfield{journal}{%
  \Doi{10.1016/j.physrep.2016.05.001}{\bibinfo {journal} {Phys. Rept.}}\ }%
  \textbf{\bibinfo {volume} {636}},\ \bibinfo {pages} {1--46} (\bibinfo {year}
  {2016}),\ \Eprint{http://arxiv.org/abs/1605.02016}{arXiv:1605.02016
  [astro-ph.HE]}%
  \bibAnnoteFile{NoStop}{Charles:2016pgz}%
\bibitem{Sirunyan:2018owy}%
  \BibitemOpen
  \bibfield{author}{%
  \bibinfo {author} {\bibfnamefont{Albert~M}\ \bibnamefont{Sirunyan}}
  \emph{et~al.} (\bibinfo {collaboration} {CMS}),\ }%
  \bibfield{title}{%
  \enquote{\bibinfo {title} {{Search for invisible decays of a Higgs boson
  produced through vector boson fusion in proton-proton collisions at $\sqrt{s}
  =$ 13 TeV}},}\ }%
  \bibfield{journal}{%
  \Doi{10.1016/j.physletb.2019.04.025}{\bibinfo {journal} {Phys. Lett.}}\ }%
  \textbf{\bibinfo {volume} {B793}},\ \bibinfo {pages} {520--551} (\bibinfo
  {year} {2019}),\ \Eprint{http://arxiv.org/abs/1809.05937}{arXiv:1809.05937
  [hep-ex]}%
  \bibAnnoteFile{NoStop}{Sirunyan:2018owy}%
\bibitem{Belanger:2020gnr}%
  \BibitemOpen
  \bibfield{author}{%
  \bibinfo {author} {\bibfnamefont{Genevieve}\ \bibnamefont{Belanger}},
  \bibinfo {author} {\bibfnamefont{Ali}\ \bibnamefont{Mjallal}},\ and\ \bibinfo
  {author} {\bibfnamefont{Alexander}\ \bibnamefont{Pukhov}},\ }%
  \bibfield{title}{%
  \enquote{\bibinfo {title} {{Recasting direct detection limits within
  micrOMEGAs and implication for non-standard Dark Matter scenarios}},}\ }%
   (\bibinfo {month} {3}\ \bibinfo {year} {2020}),\
  \Eprint{http://arxiv.org/abs/2003.08621}{arXiv:2003.08621 [hep-ph]}%
  \bibAnnoteFile{NoStop}{Belanger:2020gnr}%
\bibitem{Aghanim:2018eyx}%
  \BibitemOpen
  \bibfield{author}{%
  \bibinfo {author} {\bibfnamefont{N.}~\bibnamefont{Aghanim}} \emph{et~al.}
  (\bibinfo {collaboration} {Planck}),\ }%
  \bibfield{title}{%
  \enquote{\bibinfo {title} {{Planck 2018 results. VI. Cosmological
  parameters}},}\ }%
   (\bibinfo {year} {2018}),\
  \Eprint{http://arxiv.org/abs/1807.06209}{arXiv:1807.06209 [astro-ph.CO]}%
  \bibAnnoteFile{NoStop}{Aghanim:2018eyx}%
\bibitem{Duda:2001ae}%
  \BibitemOpen
  \bibfield{author}{%
  \bibinfo {author} {\bibfnamefont{Gintaras}\ \bibnamefont{Duda}}, \bibinfo
  {author} {\bibfnamefont{Graciela}\ \bibnamefont{Gelmini}},\ and\ \bibinfo
  {author} {\bibfnamefont{Paolo}\ \bibnamefont{Gondolo}},\ }%
  \bibfield{title}{%
  \enquote{\bibinfo {title} {{Detection of a subdominant density component of
  cold dark matter}},}\ }%
  \bibfield{journal}{%
  \Doi{10.1016/S0370-2693(02)01266-2}{\bibinfo {journal} {Phys. Lett. B}}\ }%
  \textbf{\bibinfo {volume} {529}},\ \bibinfo {pages} {187--192} (\bibinfo
  {year} {2002}),\
  \Eprint{http://arxiv.org/abs/hep-ph/0102200}{arXiv:hep-ph/0102200}%
  \bibAnnoteFile{NoStop}{Duda:2001ae}%
\bibitem{Duda:2002hf}%
  \BibitemOpen
  \bibfield{author}{%
  \bibinfo {author} {\bibfnamefont{Gintaras}\ \bibnamefont{Duda}}, \bibinfo
  {author} {\bibfnamefont{Graciela}\ \bibnamefont{Gelmini}}, \bibinfo {author}
  {\bibfnamefont{Paolo}\ \bibnamefont{Gondolo}}, \bibinfo {author}
  {\bibfnamefont{Joakim}\ \bibnamefont{Edsjo}},\ and\ \bibinfo {author}
  {\bibfnamefont{Joseph}\ \bibnamefont{Silk}},\ }%
  \bibfield{title}{%
  \enquote{\bibinfo {title} {{Indirect detection of a subdominant density
  component of cold dark matter}},}\ }%
  \bibfield{journal}{%
  \Doi{10.1103/PhysRevD.67.023505}{\bibinfo {journal} {Phys. Rev. D}}\ }%
  \textbf{\bibinfo {volume} {67}},\ \bibinfo {pages} {023505} (\bibinfo {year}
  {2003}),\ \Eprint{http://arxiv.org/abs/hep-ph/0209266}{arXiv:hep-ph/0209266}%
  \bibAnnoteFile{NoStop}{Duda:2002hf}%
\bibitem{Queiroz:2019acr}%
  \BibitemOpen
  \bibfield{author}{%
  \bibinfo {author} {\bibfnamefont{Farinaldo~S.}\ \bibnamefont{Queiroz}}\ and\
  \bibinfo {author} {\bibfnamefont{Clarissa}\ \bibnamefont{Siqueira}},\ }%
  \bibfield{title}{%
  \enquote{\bibinfo {title} {{Search for Semi-Annihilating Dark Matter with
  Fermi-LAT, H.E.S.S., Planck, and the Cherenkov Telescope Array}},}\ }%
  \bibfield{journal}{%
  \Doi{10.1088/1475-7516/2019/04/048}{\bibinfo {journal} {JCAP}}\ }%
  \textbf{\bibinfo {volume} {04}},\ \bibinfo {pages} {048} (\bibinfo {year}
  {2019}),\ \Eprint{http://arxiv.org/abs/1901.10494}{arXiv:1901.10494
  [hep-ph]}%
  \bibAnnoteFile{NoStop}{Queiroz:2019acr}%
\bibitem{Ma:2007gq}%
  \BibitemOpen
  \bibfield{author}{%
  \bibinfo {author} {\bibfnamefont{Ernest}\ \bibnamefont{Ma}},\ }%
  \bibfield{title}{%
  \enquote{\bibinfo {title} {{Z(3) Dark Matter and Two-Loop Neutrino Mass}},}\
  }%
  \bibfield{journal}{%
  \Doi{10.1016/j.physletb.2008.02.053}{\bibinfo {journal} {Phys. Lett. B}}\ }%
  \textbf{\bibinfo {volume} {662}},\ \bibinfo {pages} {49--52} (\bibinfo {year}
  {2008}),\ \Eprint{http://arxiv.org/abs/0708.3371}{arXiv:0708.3371 [hep-ph]}%
  \bibAnnoteFile{NoStop}{Ma:2007gq}%
\bibitem{Aoki:2014cja}%
  \BibitemOpen
  \bibfield{author}{%
  \bibinfo {author} {\bibfnamefont{Mayumi}\ \bibnamefont{Aoki}}\ and\ \bibinfo
  {author} {\bibfnamefont{Takashi}\ \bibnamefont{Toma}},\ }%
  \bibfield{title}{%
  \enquote{\bibinfo {title} {{Impact of semi-annihilation of $\mathbb{Z}_3$
  symmetric dark matter with radiative neutrino masses}},}\ }%
  \bibfield{journal}{%
  \Doi{10.1088/1475-7516/2014/09/016}{\bibinfo {journal} {JCAP}}\ }%
  \textbf{\bibinfo {volume} {09}},\ \bibinfo {pages} {016} (\bibinfo {year}
  {2014}),\ \Eprint{http://arxiv.org/abs/1405.5870}{arXiv:1405.5870 [hep-ph]}%
  \bibAnnoteFile{NoStop}{Aoki:2014cja}%
\bibitem{Ding:2016wbd}%
  \BibitemOpen
  \bibfield{author}{%
  \bibinfo {author} {\bibfnamefont{Ran}\ \bibnamefont{Ding}}, \bibinfo {author}
  {\bibfnamefont{Zhi-Long}\ \bibnamefont{Han}}, \bibinfo {author}
  {\bibfnamefont{Yi}~\bibnamefont{Liao}},\ and\ \bibinfo {author}
  {\bibfnamefont{Wan-Peng}\ \bibnamefont{Xie}},\ }%
  \bibfield{title}{%
  \enquote{\bibinfo {title} {{Radiative neutrino mass with $\mathbb{Z}_{3}$
  dark matter: from relic density to LHC signatures}},}\ }%
  \bibfield{journal}{%
  \Doi{10.1007/JHEP05(2016)030}{\bibinfo {journal} {JHEP}}\ }%
  \textbf{\bibinfo {volume} {05}},\ \bibinfo {pages} {030} (\bibinfo {year}
  {2016}),\ \Eprint{http://arxiv.org/abs/1601.06355}{arXiv:1601.06355
  [hep-ph]}%
  \bibAnnoteFile{NoStop}{Ding:2016wbd}%
\bibitem{Ho:2017fte}%
  \BibitemOpen
  \bibfield{author}{%
  \bibinfo {author} {\bibfnamefont{Shu-Yu}\ \bibnamefont{Ho}}, \bibinfo
  {author} {\bibfnamefont{Takashi}\ \bibnamefont{Toma}},\ and\ \bibinfo
  {author} {\bibfnamefont{Koji}\ \bibnamefont{Tsumura}},\ }%
  \bibfield{title}{%
  \enquote{\bibinfo {title} {{A Radiative Neutrino Mass Model with SIMP Dark
  Matter}},}\ }%
  \bibfield{journal}{%
  \Doi{10.1007/JHEP07(2017)101}{\bibinfo {journal} {JHEP}}\ }%
  \textbf{\bibinfo {volume} {07}},\ \bibinfo {pages} {101} (\bibinfo {year}
  {2017}),\ \Eprint{http://arxiv.org/abs/1705.00592}{arXiv:1705.00592
  [hep-ph]}%
  \bibAnnoteFile{NoStop}{Ho:2017fte}%
\bibitem{Kannike:2012pe}%
  \BibitemOpen
  \bibfield{author}{%
  \bibinfo {author} {\bibfnamefont{Kristjan}\ \bibnamefont{Kannike}},\ }%
  \bibfield{title}{%
  \enquote{\bibinfo {title} {{Vacuum Stability Conditions From Copositivity
  Criteria}},}\ }%
  \bibfield{journal}{%
  \Doi{10.1140/epjc/s10052-012-2093-z}{\bibinfo {journal} {Eur. Phys. J. C}}\
  }%
  \textbf{\bibinfo {volume} {72}},\ \bibinfo {pages} {2093} (\bibinfo {year}
  {2012}),\ \Eprint{http://arxiv.org/abs/1205.3781}{arXiv:1205.3781 [hep-ph]}%
  \bibAnnoteFile{NoStop}{Kannike:2012pe}%
\bibitem{Kannike:2016fmd}%
  \BibitemOpen
  \bibfield{author}{%
  \bibinfo {author} {\bibfnamefont{Kristjan}\ \bibnamefont{Kannike}},\ }%
  \bibfield{title}{%
  \enquote{\bibinfo {title} {{Vacuum Stability of a General Scalar Potential of
  a Few Fields}},}\ }%
  \bibfield{journal}{%
  \Doi{10.1140/epjc/s10052-016-4160-3}{\bibinfo {journal} {Eur. Phys. J. C}}\
  }%
  \textbf{\bibinfo {volume} {76}},\ \bibinfo {pages} {324} (\bibinfo {year}
  {2016}),\ \bibinfo {note} {[Erratum: Eur.Phys.J.C 78, 355 (2018)]},\
  \Eprint{http://arxiv.org/abs/1603.02680}{arXiv:1603.02680 [hep-ph]}%
  \bibAnnoteFile{NoStop}{Kannike:2016fmd}%
\bibitem{Staub:2008uz}%
  \BibitemOpen
  \bibfield{author}{%
  \bibinfo {author} {\bibfnamefont{F.}~\bibnamefont{Staub}},\ }%
  \bibfield{title}{%
  \enquote{\bibinfo {title} {{SARAH}},}\ }%
   (\bibinfo {month} {6}\ \bibinfo {year} {2008}),\
  \Eprint{http://arxiv.org/abs/0806.0538}{arXiv:0806.0538 [hep-ph]}%
  \bibAnnoteFile{NoStop}{Staub:2008uz}%
\bibitem{Staub:2013tta}%
  \BibitemOpen
  \bibfield{author}{%
  \bibinfo {author} {\bibfnamefont{Florian}\ \bibnamefont{Staub}},\ }%
  \bibfield{title}{%
  \enquote{\bibinfo {title} {{SARAH 4 : A tool for (not only SUSY) model
  builders}},}\ }%
  \bibfield{journal}{%
  \Doi{10.1016/j.cpc.2014.02.018}{\bibinfo {journal} {Comput. Phys. Commun.}}\
  }%
  \textbf{\bibinfo {volume} {185}},\ \bibinfo {pages} {1773--1790} (\bibinfo
  {year} {2014}),\ \Eprint{http://arxiv.org/abs/1309.7223}{arXiv:1309.7223
  [hep-ph]}%
  \bibAnnoteFile{NoStop}{Staub:2013tta}%
\end{thebibliography}%

\end{document}